\documentclass{emulateapj}

\shorttitle{Luminous high-redshift Quasar Survey with SDSS-{\it WISE} }
\shortauthors{Wang et al.}

\begin{document}

\title{A survey of luminous high-redshift quasars with SDSS and {\it WISE}. I. target selection and optical spectroscopy}

\author{Feige Wang\altaffilmark{1,2}, Xue-Bing Wu\altaffilmark{1,3}, Xiaohui Fan\altaffilmark{2,3}, Jinyi Yang\altaffilmark{1,2}, Weimin Yi\altaffilmark{4,5}, 
            Fuyan Bian\altaffilmark{6,10}, Ian D. McGreer\altaffilmark{2}, Qian Yang\altaffilmark{1,2}, Yanli Ai\altaffilmark{7}, Xiaoyi Dong\altaffilmark{1}, 
            Wenwen Zuo\altaffilmark{8},  Linhua Jiang\altaffilmark{3}, Richard Green\altaffilmark{2}, Shu Wang\altaffilmark{1}, Zheng Cai\altaffilmark{2,9}, 
            Ran Wang\altaffilmark{3}, Minghao Yue\altaffilmark{1}}

\altaffiltext{1}{Department of Astronomy, School of Physics, Peking University, Beijing 100871, China}
\altaffiltext{2}{Steward Observatory, University of Arizona, 933 North Cherry Avenue, Tucson, AZ 85721, USA}
\altaffiltext{3}{Kavli Institute for Astronomy and Astrophysics, Peking University, Beijing 100871, China}
\altaffiltext{4}{Yunnan Observatories, Chinese Academy of Sciences, Kunming 650011,China}
\altaffiltext{5}{Key Laboratory for the Structure and Evolution of Celestial Objects, Chinese Academy of Sciences, Kunming 650011,China}
\altaffiltext{6}{Research School of Astronomy and Astrophysics, Australian National University, Weston Creek, ACT 2611, Australia}
\altaffiltext{7}{School of Physics and Astronomy, Sun Yat-Sen University, Guangzhou 510275, China}
\altaffiltext{8}{Shanghai Astronomical Observatory, Chinese Academy of Sciences, Shanghai 200030, China}
\altaffiltext{9}{UCO/Lick Observatory, University of California, 1156 High Street, Santa Cruz, CA 95064, USA}
\altaffiltext{10}{Stromlo Fellow}


\begin{abstract}
High-redshift quasars are important tracers of structure and evolution in the early universe. However, they are very rare and difficult to find when using color selection because of contamination from late-type dwarfs. High-redshift quasar surveys based on only optical colors suffer from incompleteness and low identification efficiency, especially at $z\gtrsim4.5$. 
We have developed a new method to select $4.7\lesssim z \lesssim 5.4$ quasars with both high efficiency and completeness by combining optical and mid-IR {\it Wide-field Infrared Survey Explorer (WISE)} photometric data, and are conducting a luminous $z\sim5$ quasar survey in the whole Sloan Digital Sky Survey (SDSS) footprint. We have spectroscopically observed 99 out of 110 candidates with $z$-band magnitudes brighter than 19.5 and 64 (64.6\%) of them are quasars with redshifts of $4.4\lesssim z \lesssim 5.5$ and absolute magnitudes of $-29\lesssim M_{1450} \lesssim -26.4$. In addition, we also observed 14 fainter candidates selected with the same criteria and identified 8 (57.1\%) of them as quasars with $4.7<z<5.4$ .
Among 72 newly identified quasars, 12 of them are at $5.2 < z < 5.7$, which leads to an increase of 
$\sim$36\% of the number of known quasars at this redshift range. More importantly, our identifications doubled the number of quasars with $M_{1450}<-27.5$ at $z>4.5$, which will set strong constraints on the bright end of the quasar luminosity function.  We also expand our method to select quasars at $z\gtrsim5.7$. In this paper we report the discovery of  four new luminous $z\gtrsim5.7$ quasars based on SDSS-{\it WISE} selection. 

\end{abstract}


\section{Introduction}
As the most luminous non-transient objects that can be observed in the early universe, high-redshift quasars are important tracers to study early structure formation and the history of cosmic reionization \citep[e.g.][]{fan06a}. In addition, understanding the evolution of quasars from the early universe to the present epoch allows us to study the accretion history of supermassive black holes (SMBHs). However,  high-redshift quasar searches are highly challenging due to their low spatial density and a high rate of contamination from cool dwarfs when using the traditional multicolor selection method. 

With the increasing number of large surveys such as the 2dF Quasar Redshift Survey \citep[2QZ;][]{croom01} and the Sloan Digital Sky Survey \citep[SDSS;][]{york00}, the number of known quasars has been increasing rapidly. The 2QZ identified more than 23,000 $B<21$ quasars \citep{croom04}. The first two phases of the SDSS spectroscopically identified more than 100,000 quasars \citep{schneider10}, and the Baryonic Oscillation Spectroscopic Survey \citep[BOSS;][]{dawson13}, which is the third phase of SDSS \citep[SDSS-III;][]{eisenstein11}, spectroscopically identified more than 300,000 quasars \citep[e.g.][]{paris12,paris14} selected by using the extreme deconvolution method \citep{bovy11,dipompeo15}. However, most of these quasars are selected based on optical colors only and mostly at lower redshift ($z\lesssim$3.5). Based on SDSS $g-r$/$r-i$ and $r-i$/$i-z$ colors, several hundred $z\ge4$ quasars and some $z\ge5$ quasars have been discovered \citep{fan99,fan00a,fan01b,zheng00,anderson01,schneider01,chiu05}. These quasar surveys have to make a very strict $r-i$/$i-z$ cut and suffer from low completeness to avoid the strong contamination of late-type stars. Nonetheless, the success rate of finding $z\gtrsim 4.5$ quasars in automated spectroscopic surveys remains quite low. For example, the overall success rate of finding $z\gtrsim 4.5$ quasars in the SDSS quasar survey is less than 10\%. 

\cite{cool06} discovered three $z>5$ quasars in the AGN and Galaxy Evolution Survey \citep[AGES;][]{kochanek12}, with targets selected from {\it Spitzer Space Telescope} mid-infrared photometry. The combination of optical and near-IR colors can improve the success rate and completeness of selecting high-redshift quasars. \cite{mcgreer13} identified 73 $4.7\le z \le 5.1$ quasars out of 92 candidates by adding near-IR $J$-band photometry. However, this method can be applied only to a narrow redshift range \citep{mcgreer13}.

Spectroscopic followup of SDSS $i$-dropout objects has identified more than 30 luminous $z>5.7$ quasars \citep{fan00b,fan01a,fan03,fan04,fan06b,jiang08,jiang09,jiang15}. The Canada$-$France High-$z$ Quasar Survey (CFHQS) has found 20 fainter $z>5.7$ quasars based on multicolor optical imaging at the Canada$-$France$-$Hawaii Telescope (CFHT) \citep{willott07,willott09,willott10a,willott10b}. The Panoramic Survey Telescope \& Rapid Response System 1 \citep[Pan-STARRS1, PS1][]{kaiser02,kaiser10} high-redshift quasar survey has discovered more than ten $z>5.7$ quasars \citep{morganson12,banados14}. 
Recently, \cite{banados15} improved the efficiency for selecting $z\sim6$ quasars by matching Pan-STARRS1 and Faint Images of the Radio Sky at Twenty Centimeters \citep[FIRST,][]{becker95} and \cite{carnall15} obtained a cleaner $z\gtrsim5.7$ candidate sample by matching optical photometry from the Very Large Telescope Survey Telescope ATLAS survey \citep{shanks15} and {\it Wide-field Infrared Survey Explorer (WISE)} photometry.
However,  these methods can be used to select only $z<6.5$ quasars. The first $z>6.5$ quasar, ULAS J112001.48$+$064124.3 at $z = 7.1$, was discovered in the United Kingdom Infrared Deep Sky Survey (UKIDSS) Large Area Survey \citep[LAS;][]{lawrence07} by \cite{mortlock11}. Recently, six more $z>6.5$ quasars \citep{venemans13,venemans15} were discovered in the Visible and Infrared Survey Telescope for Astronomy (VISTA) Kilo-Degree Infrared Galaxy survey \citep[VIKING;][]{arnaboldi07} and Pan-STARRS1 survey. 
To date, although more than 300,000 quasars are known, among them are about 170 quasars at $z>5$,  $\sim$60 quasars at $z>6$ and one quasar at $z>7$.  In addition there is an obvious gap of known quasars at $5.2\lesssim z \lesssim 5.7$, which is caused by their optical colors being very similar to those of late-type stars, especially M dwarfs. This redshift distribution gap of known quasars poses challenges for the studies of the high-redshift  quasar luminosity function (QLF), the black hole mass function (BHMF) and the properties of the post-reionziation intergalactic medium (IGM).

NASA's {\it WISE} \citep{wright10} mapped the sky at 3.4, 4.6, 12, and 22 $\mu$m ($W1$, $W2$, $W3$, $W4$) with an angular resolution of 6.1, 6.4, 6.5, and 12.0 arcsec and 5$\sigma$ photometric sensitivity better than 0.08, 0.11, 1, and 6 mJy (corresponding to 16.5, 15.5, 11.2, and 7.9 Vega magnitudes) in these four bands, respectively. The {\it WISE} All-Sky Data Release\footnote{http://wise2.ipac.caltech.edu/docs/release/allsky/} includes all data taken during the {\it WISE} full cryogenic mission phase from 2010 January 7 to August 6 and consists of over 563 million objects. Recently the ALLWISE\footnote{http://wise2.ipac.caltech.edu/docs/release/allwise/} program combined data from the {\it WISE} cryogenic and NEOWISE \citep{mainzer11} post-cryogenic survey phases to form the most comprehensive view of the full mid-infrared sky. The ALLWISE photometric catalog includes over 747 million objects with enhanced photometric sensitivity and accuracy and improved astrometric precision compared to the {\it WISE} All-Sky Data Release. 

In this paper we present  a new robust method for selecting luminous high-redshift quasars by combining ALLWISE and SDSS photometric data and provide the results of optical spectroscopy followup observations. This paper is organized as follows. In \textsection 2 we summarize the ALLWISE detection rate of high-redshift quasars and the $W1-W2$ colors of high-redshift quasars and late-type stars. In \textsection 3 we describe our target selection and spectroscopic observations. In \textsection 4 we present our spectroscopic quasar sample and in \textsection 5 we discuss the pros and cons of our selection method and compare with the SDSS high-redshift quasar selection method. In \textsection 6 we extend our selection method to $z\gtrsim5.7$ quasars and present the discovery of four new quasars at $z>5.7$, and in \textsection 7 we give a brief summary. 
Throughout the paper, SDSS magnitudes are reported on the asinh system \citep{lupton99}, and {\it WISE} magnitudes are on the Vega
system. We adopt a standard $\Lambda$CDM cosmology with Hubble constant $H_0=70~{\rm km~s}^{-1}~{\rm Mpc}^{-1}$, and density parameters $\Omega_M=0.3$ and $\Omega_{\Lambda}=0.7$.

\section{WISE Photometry of Published High-Redshift Quasars}

\begin{figure}
\includegraphics[width=0.5\textwidth]{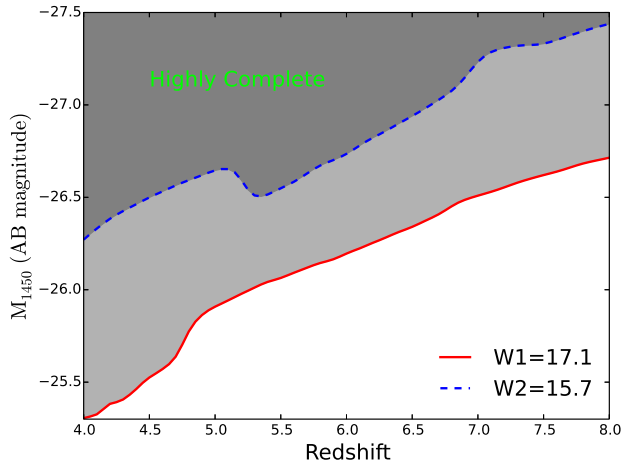}
\caption{$M_{1450}$ as a function of redshift for a type I quasar template from \cite{glikman06}. The red solid line denotes $W1=17.1$ and the blue dashed line denotes $W2=15.7$, which are the 95\% completeness limits of the ALLWISE catalog.  ALLWISE is a promising data set for finding  high-redshift quasars with high luminosity (e.g. $M_{1450}\lesssim-26.5$ at $z\lesssim6$ and $M_{1450}\lesssim-27.0$ at $z\lesssim7$).
\label{fig1}}
\end{figure}

\cite{wu12} first studied SDSS quasars in the {\it WISE} preliminary data release \footnote{http://wise2.ipac.caltech.edu/docs/release/prelim/} sky coverage and found that {\it WISE} can detect more than 50\% of the SDSS quasars with $i<20.5$ and $W1-W2>0.57$ and can separate late-type stars and quasars efficiently.  Remarkably, this method has been used in the Large Sky Area Multi-Object Fiber Spectroscopic Telescope (LAMOST) Quasar Survey and has discovered several thousand new quasars mainly at $z\lesssim 4.0$ \citep{ai15}.
The depth of ALLWISE has improved from the early catalogs, due to  stacking of multiple epoch photometry; and the 95\% completeness limits of ALLWISE $W1$ and $W2$ are at about 17.1 (44 $\mu$Jy) and 15.7 (88 $\mu$Jy) magnitude.\footnote{http://wise2.ipac.caltech.edu/docs/release/allwise/expsup/}.

Figure 1 shows the absolute magnitudes at $1450\AA$ of a low redshift type I quasar template \citep{glikman06} as a function of redshift at the ALLWISE $W1$ and $W2$ magnitude limits. It is clear that the ALLWISE data set has a high completeness of detecting luminous high-redshift quasars (e.g. $M_{1450}\lesssim-26.5$ at $z\lesssim6$ and $M_{1450}\lesssim-27.0$ at $z\lesssim7$), and is a highly valuable data set for finding luminous high-redshift quasars when combined with other optical sky surveys such as SDSS, PanSTARRS, SkyMapper, DES, VST ATLAS, DECaLS, and LSST as well as with near-IR surveys, such as UKIDSS-LAS, UKIDSS-UHS and VISTA-VHS. 

\begin{figure}
\includegraphics[angle=0,scale=0.38]{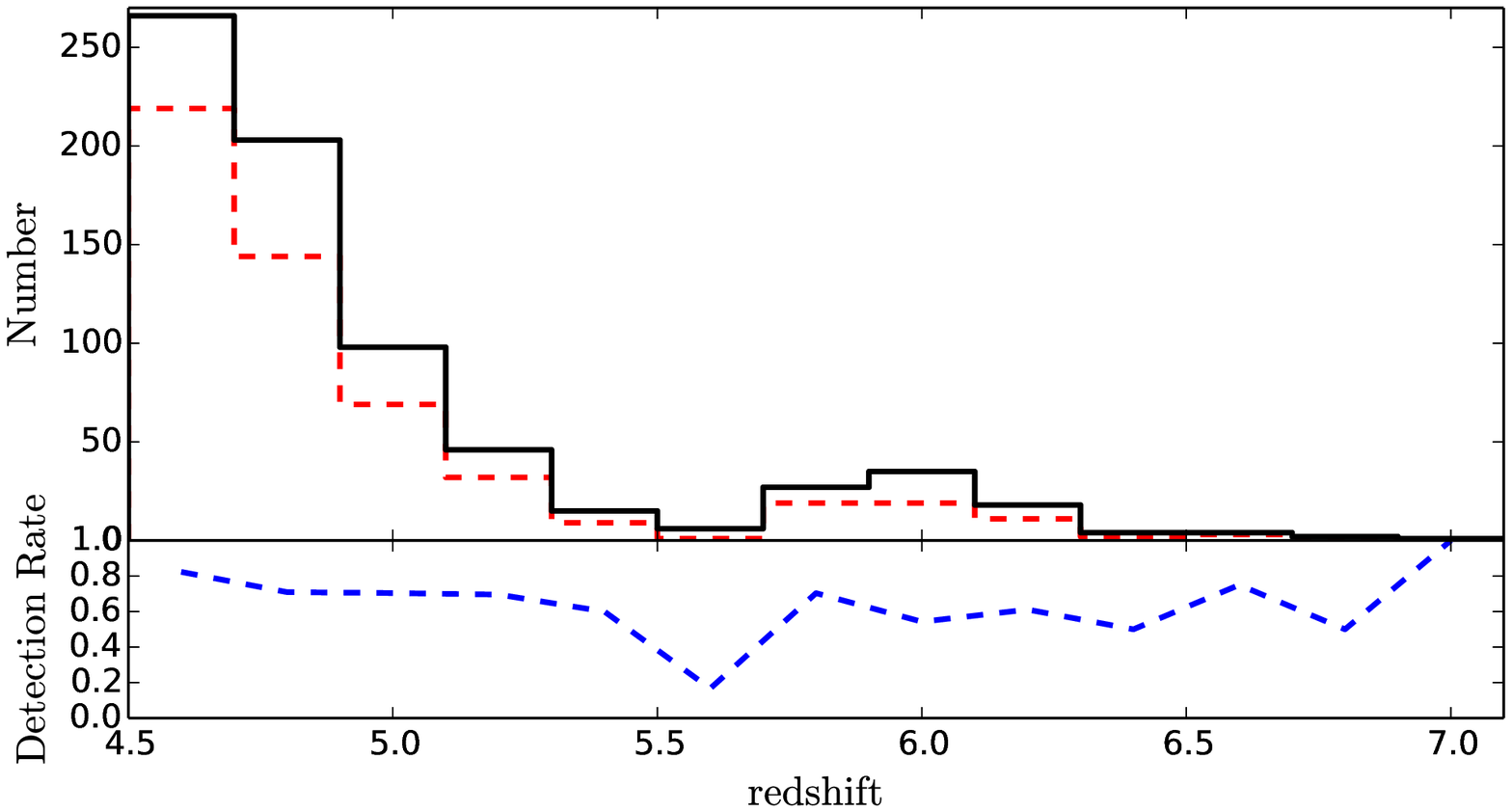}
\includegraphics[angle=0,scale=0.38]{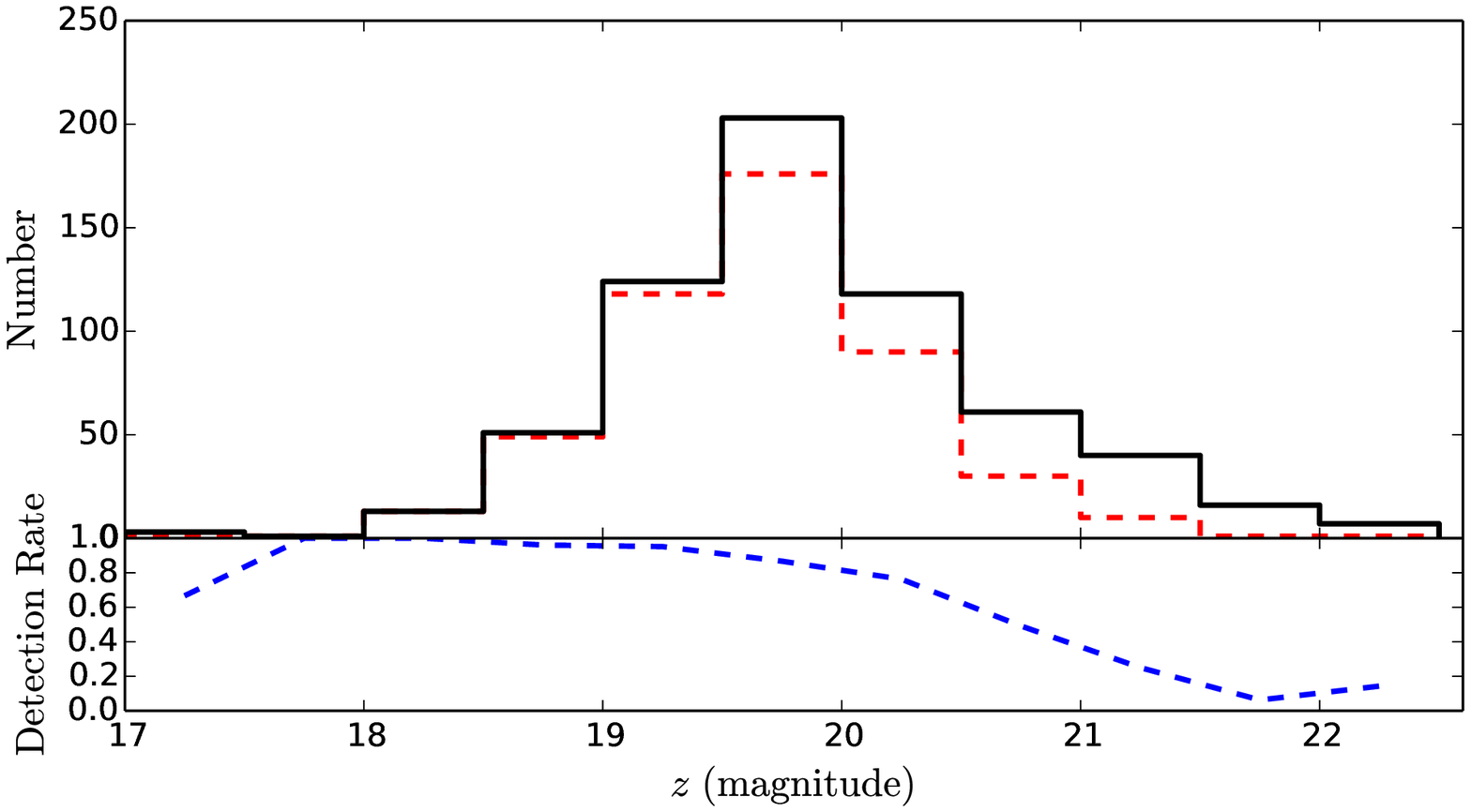}
\caption{Upper panel: the black solid line denotes the redshift histogram of all published $z>4.5$ quasars, while the red dashed line is the redshift histogram of the ALLWISE-detected $z>4.5$ quasars. The ratio between them (blue dashed line) is also plotted as a function of redshift. Note that the drop of the detection rate at $z\sim5.5$ is mainly affected by the statistic error because of the small number of objects in the bin. Lower panel: the black solid line denotes the SDSS $z$-band magnitude histogram of $z\ge4.5$ quasars with SDSS detections, while the red dashed line is the $z$-band magnitude histogram of the ALLWISE-detected $z\ge4.5$ quasars. The ratio between them is also plotted as a function of magnitude.
\label{fig2}}
\end{figure}

We collected 725 published quasars at $z>4.5$ from the SDSS quasar catalogs and the literature (Table 1). We cross-matched these high-redshift quasars with the ALLWISE source catalog using a position offset within $2''$, and found that 530 (73.1\%) of them were detected in the ALLWISE $W1$ band, 505 (69.7\%) in the $W2$ band, 261 (36.0\%) in the $W3$ band, and 94 (13.0\%) in the $W4$ band. In Figure 2, we show the redshift and $z$-band magnitude distribution of these quasars as well as the dependence of the ALLWISE detection rate on the redshift and magnitude. 
The ALLWISE data set detected $\gtrsim$50\% of the known quasars over almost all redshifts and  $\gtrsim95$\% of the known quasars with $z\lesssim19.5$. This is consistent with what we find in Figure 1. In particular, 34 of 50 (68\%) published $z\ge6$ quasars are detected by the ALLWISE data set, which is higher than the result reported in previous studies on the {\it WISE} All-Sky detection rate (17/31, 55\%) of $z>6$ quasars based on a smaller sample \citep{blain13}. Note that the drop in the detection rate at the brightest end is caused by the fact that one of three quasars in the brightest bin is blended by a nearby bright star in the ALLWISE image. 

\begin{deluxetable*}{ccrrrrrrrrrrrrrrcrl}
\tabletypesize{\scriptsize}
\tablecaption{Optical and {\it WISE} Photometry of  725 published $z>4.5$ quasars. \label{tbl-1}}
\tablewidth{0pt}
\tablehead{
\colhead{Name} & \colhead{Redshift} &\colhead{Ref} &
\colhead{$r$} &\colhead{$\sigma_{r}$}&\colhead{$i$} &\colhead{$\sigma_{i}$}& \colhead{$z$} &\colhead{$\sigma_{z}$}& \colhead{Opt} &
 \colhead{$W1$} &\colhead{$\sigma_{W1}$}& \colhead{$W2$} &\colhead{$\sigma_{W2}$} & \colhead{$WISE$}
}
\startdata
J000239.39$+$255034.80&5.800 &16&23.09 &0.32 &21.51 &0.11 &18.96 &0.05 &DR10&16.16 &0.06 &15.54 &0.13 &AW \\
J000552.34$-$000655.80&5.850 &16&24.97 &0.51 &22.98 &0.28 &20.41 &0.13 &DR10&17.30 &0.16 &17.04 &99.0 &AW \\
J000651.61$-$620803.70&4.510 &51&18.29 &99.0 &99.0 &99.0 &99.0 &99.0 &REF &15.20 &0.03 &14.61 &0.04 &AW \\
J000749.17$+$004119.61&4.780 &DR12&21.36 &0.06 &19.97 &0.03 &19.84 &0.08 &DR10&16.85 &0.11 &16.41 &0.26 &AW \\
J000825.77$-$062604.60&5.929 &24&23.91 &0.59 &23.55 &0.63 &20.01 &0.14 &DR10&16.81 &0.11 &15.68 &0.14 &AW \\
J001115.24$+$144601.80&4.964 &DR12&19.48 &0.02 &18.17 &0.02 &18.03 &0.03 &DR10&15.29 &0.04 &14.69 &0.06 &AW \\
J001207.79$+$094720.23&4.745 &DR12&21.40 &0.07 &19.81 &0.04 &19.86 &0.10 &DR10&16.26 &0.07 &15.80 &0.17 &AW \\
J001529.86$-$004904.30&4.930 &32&22.59 &0.15 &20.99 &0.05 &20.56 &0.12 &DR10&99.0 &99.0 &99.0 &99.0 &99 \\
J001714.68$-$100055.43&5.011 &DR7&21.23 &0.06 &19.45 &0.03 &19.55 &0.09 &DR10&15.94 &0.06 &15.17 &0.09 &AW \\
J002208.00$-$150539.76&4.528 &50&19.39 &0.03 &18.75 &0.02 &18.54 &0.04 &DR10&15.54 &0.05 &15.11 &0.10 &AW 
\enddata
\tablecomments{Table \ref{tbl-1} is available in its entirety in the electronic edition of the journal. The first 10 rows are shown here for guidance regarding its form and content. The names here are in the format of JHHMMSS.SS$+$/$-$DDMMSS.SS. The Ref column lists the reference for each quasar and the Opt column lists the references for the optical magnitudes. Most optical magnitudes are from the SDSS DR10 photometric catalog and are Galactic extinction corrected SDSS PSF asinh magnitudes ($z_{AB} = z_{SDSS} + 0.02$ mag).  
 The optical magnitudes come from the reference paper and are in the AB system if the quasar does not have SDSS DR 10 photometry. The last column lists the flag from the {\it WISE} data: AW = ALLWISE catalog, AWR = ALLWISE Reject catalog, AS = ALL-SKY {\it WISE} catalog, ASR = ALL-SKY {\it WISE} Reject catalog, and 99 means no detection in any {\it WISE} catalog. This table only includes quasars that were published before 2015 July.}

\end{deluxetable*}

\begin{figure}
\includegraphics[width=0.5\textwidth]{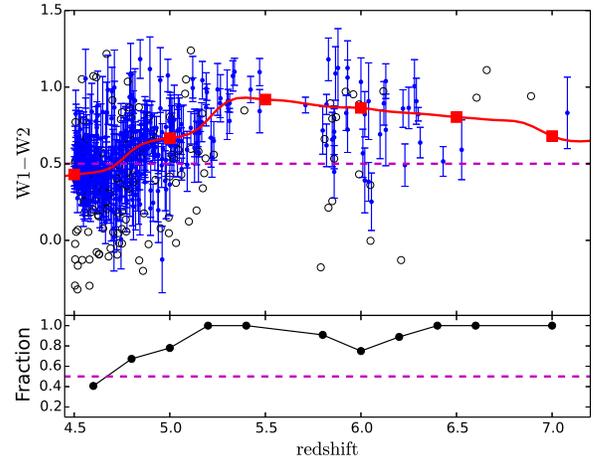}
\caption{Upper: the $W1-W2$ vs. redshift diagram. The purple dashed line represents W1$-$W2$=0.5$. The red solid line represents the color-$z$ relation predicted using the quasar template from \cite{glikman06}. The solid squares from left to right mark the color tracks for quasars from $z=5.0$ to $z=7.0$ in steps of $\Delta z =0.5$. The blue points represent $z\ge4.5$ quasars detected by ALLWISE with both $W1$ and $W2$ at five sigma level. The black open cycles denote $z\ge4.5$ quasars detected by ALLWISE with only W1 at five sigma level.
Lower: the fraction of $z>4.5$ quasars with $W1-W2>0.5$ as a function of redshift. Note that we only consider quasars detected by ALLWISE with both W1 and W2 at five sigma level here. The purple dashed line denotes a fraction of 50\%. We calculated the fraction using a redshift bin of 0.2. Note that the bins at $z\sim5.6$ and $z\sim6.8$ are not plotted due to no available data. The drop of the fraction at $z\sim6$ is partly because  of the large uncertainties of {\it WISE} photometry for faint high-redshift quasars.}
\label{fig3}
\end{figure}

Due to their red optical colors,  late-type stars  (especially M dwarfs) are the most serious contaminants in selecting $z\sim5$ quasar candidates using optical colors (e.g. Fan et al. 1999). Wu et al. (2012) studied the color distributions of {\it WISE}-detected quasars and stars with SDSS spectroscopy and found that both normal and late-type stars can be well rejected with $W1-W2 > 0.57$ (see Figure 6 in Wu et al. 2012). Figure 3 shows the $W1-W2$ color of quasars as a function of redshift. To have a clear view of the color-redshift track, we plot quasars with both $W1$ and $W2$ having a signal-to-noise ratio great than five as blue points and quasars with only $W1$ at the five sigma level as open black cycles in Figure 3. In order to get a more reasonable statistics of known quasars, we only count quasars detected by ALLWISE with both W1 and W2 at five sigma level in the lower panel in Figure 3.
Clearly most high-redshift quasars, especially $z>4.7$ quasars, have a red W1$-$W2 color. There are $\sim41$\% of $z<4.7$ quasars having red $W1-W2$ colors with $W1-W2>0.5$, $\sim68$\% of $4.7<z<5.0$ quasars with $W1-W2>0.5$,  and $\sim$92\% of $z>5.0$ quasars with $W1-W2>0.5$.
Because {\it WISE} data have a high detection rate of luminous high-redshift quasars and can provide an effective way of separating quasars and late-type stars, we are conducting a luminous quasar survey at $z\gtrsim4.7$ by combing SDSS and ALLWISE photometry. Note that there is a glaring gap of known quasars at $5.2<z<5.7$, which can be seen in both Figures 1 and 3. This is because quasar colors are very similar to those of M dwarfs and hard to distinguish in both optical and near-IR wavelengths. Benefiting from the different $W1-W2$ colors of quasars and M dwarfs, we expect that we can select $z>5.2$ quasars more effectively by combining SDSS and ALLWISE photometry than previous quasar selection methods.

\section{Target Selection and Spectroscopic Observations}

\subsection{Target Selection}

At $z\sim5$, most quasars are undetectable in $u$-band and $g$-band because of the presence of Lyman limit systems (LLSs), which are optically thick to the continuum radiation from quasars \citep{fan99}. 
Meanwhile the Lyman series line absorptions and Lyman continuum absorptions begin to dominate in the $r$-band and  $\rm Ly\alpha$ emission moves to the $i$-band. The $r-i/i-z$ color$-$color diagram was often used to select $z\sim5$ quasar candidates in previous studies \citep{fan99,richards02,mcgreer13}. In Figure 4, we show the $r-i/i-z$ colors of stars and 274 SDSS and BOSS $4.7\le z < 5.6$ quasars \citep{schneider10,paris14}, as well as different $z\sim5$ quasar selection criteria \citep{richards02,mcgreer13}. The optical color selection limits shown here (cyan and orange dashed lines) are effective for quasars at $z\lesssim5.1$, but the selection becomes very incomplete for quasars at  $z\gtrsim 5.1$, when they enter the M star locus on the $r-i/i-z$ color$-$color diagram \citep{richards02,mcgreer13}. 
This is consistent with the color$-$redshift tracks (green solid line) derived from the $z\sim5$ quasar composite spectrum constructed from BOSS quasar spectra using a median algorithm by us.
As we discussed in \S2, $W1-W2$ can be used to reject M type stars effectively and the addition of {\it WISE} photometry will allow us to loosen the typical $r-i/i-z$ cuts to reach higher redshift while still being able to reject most late-type star contaminants. 
Following are our selection criteria.

\begin{figure}
\includegraphics[width=0.5\textwidth]{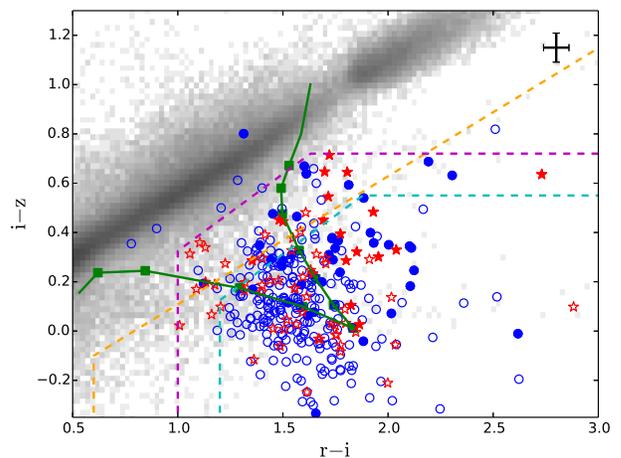}
\caption{The $i-z$ vs. $r-i$ color$-$color diagram. The purple dashed line represents our selection criteria for quasar candidates. The orange dashed line represents the SDSS $z>4.5$ quasar selection criteria \citep{richards02} and the cyan dashed line denotes $4.7\lesssim z\lesssim 5.1$ quasar selection criteria \citep{mcgreer13}. The green solid line represents the color$-$z relation predicted using $z \sim 5.0$ SDSS quasar composite spectra. The solid squares mark the color tracks for quasars from $z=4.4$ to $z=5.4$ in steps of $\Delta z =0.1$. The gray map denotes SDSS stars, the blue open cycles denote SDSS $4.7\le z < 5.1$ quasars the blue solid circles denote SDSS $5.1\le z <5.6$ quasars. The red open stars denote our newly discovered quasars with $z <5.1$, and the red solid stars denote our newly discovered quasars with $5.1\le z <5.6$. Typical error bars are shown in the upper right-hand corner.
\label{fig4}}
\end{figure}

\begin{equation}
u>22.3, z<20.2
\end{equation}
\begin{equation}
g>24.0~ or ~ g-r>1.8
\end{equation}
\begin{equation}
r-i > 1.0
\end{equation}
\begin{equation}
i-z < 0.72
\end{equation}
\begin{equation}
i-z < 0.625\times(r-i)-0.3
\end{equation}
\begin{equation}
z-W1 > 2.5
\end{equation}
\begin{equation}
W1-W2 > 0.5
\end{equation}
\begin{equation}
W1 < 17.0, \sigma_{W2} < 0.2
\end{equation}
\begin{equation}
z-W1 > 2.8~ or~ W1-W2 > 0.7, if~ i-z>0.4
\end{equation}
where optical magnitudes are Galactic extinction corrected SDSS point-spread function asinh magnitudes and the $W1$ and $W2$ magnitudes are Vega-based magnitudes.
The $u$-band and $g$-band cuts are the typical magnitude limits for dropout bands (Fan et al. 1999). The $z$-band magnitude cut is to ensure the accuracy of the $z$-band photometry since the $5\sigma$ detection of the SDSS $z$-band for point sources with $1''$ image quality is about 20.5. 
The spectral energy distributions of $z\gtrsim4.5$ quasars are mainly dominated by a power-law spectrum with a slope around $\alpha_{\nu} \sim -0.5$ \citep{vanden01}, which is flatter than that of M dwarfs. This difference leads to a redder $z-$W1 color for quasars than for M dwarfs (Eq. (6)). As we discussed in the last section, the $W1-W2$ color can separate quasars and late-type stars very efficiently; here we require $W1-W2>0.5$ (Eq. (7)). We use the magnitude or photometric error cuts of the ALLWISE photometric data (Eq. (8)) to ensure the accuracy of the $W1-W2$ color. Considering the serious contamination and redder $W1-W2$ colors for $z\gtrsim5.2$ quasars (Figure 3), we also require a more strict $z-W1$ and $W1-W2$ color accuracy for candidates with $i-z>0.4$ (Eq. (9)). Although using Eq. (9) leads to a lower completeness of selecting $z\gtrsim5.2$ quasars, it helps to reduce star contamination significantly. 
The color distribution of quasars in the color$-$color diagrams are broader than that derived from composite quasar spectra. This is not only because of the magnitude uncertainties but also the broad distributions of quasar emission line strength and continuum slopes.
The $z$-band covers the rest-frame UV continuum ($\sim$1300-1700$\rm \AA$), but the $W1$ band covers the rest-frame optical-continuum ($\sim$5000-6500$\rm \AA$). There is a break of the continuum slope at around $5500\AA$ \citep{vanden01,glikman06} and the distributions of the continuum slopes at UV and optical wavelengths are broad \citep[e.g.][]{shen11}.
In addition, as the $\rm H \alpha$ emission contributes significantly to the flux in {\it WISE} $W1$ and $W2$ bands, the different strength of the $\rm H \alpha$ emission will also cause some scatters of the $z-$W1 colors. So the large scatter of $z-W1$ colors is not only affected by the {\it WISE} magnitude uncertainties but also by the broad distributions of the UV and optical-continuum slopes and the strength of the emission lines.

The purple dashed lines in Figure 4 and Figure 5 denote our color$-$color selection criteria (Eq. (5-7)). Apparently, our $r-i/i-z$ selection criteria are much looser than those of other studies (the region between the purple dashed line and the orange dashed line) which will improve the completeness of $z\gtrsim5.1$ quasars. Among 274 SDSS and BOSS $4.7\le z < 5.6$ quasars 22 of them (blue crosses between the purple dashed line and the orange dashed line in Figure 4) satisfy our $r-i/i-z$ cuts but not the cuts in \cite{richards02}, and seven ($\sim 32$\%) of them with $z>5.1$. Except for the one that was not detected by ALLWISE, the other six $z>5.1$ quasars also satisfy our $z-W1 / W1-W2$ selection criteria shown in Figure 5. Therefore we can expect to improve the completeness of selecting $z\gtrsim5.1$ quasars with our method by combining SDSS and ALLWISE.

\begin{figure}
\includegraphics[width=0.5\textwidth]{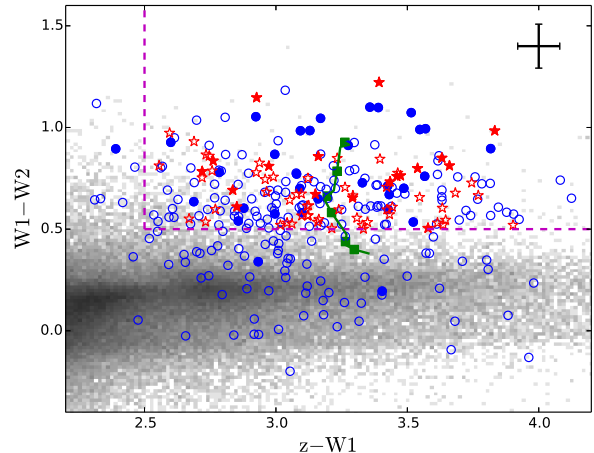}
\caption{The $z-W1$ vs. $W1-W2$ color$-$color diagram. The purple dashed line represents our selection criteria for quasar candidates. The green solid line represents the color-z relation predicted using quasar composite spectra \citep{glikman06}. The solid squares mark the color tracks for quasars from $z=4.4$ to $z=5.4$, in steps of $\Delta z =0.2$. Other symbols have the same meaning as in Fig. 4.
 \label{fig5}}
\end{figure}

We started our quasar candidate selection from a catalog of SDSS Data Release 10 (DR10) primary point sources. Applying  the optical magnitude and color cuts (Eqs (1)-(5)) using SDSS-III DR10 Query/CasJobs\footnote{http://skyserver.sdss.org/casjobs/} results in 457,930 point sources. We then cross-identified these sources with the ALLWISE source catalog using a position offset within $2''$; this reduced our candidate list to 80,404 sources with ALLWISE detections. We selected our candidates using Eqs (6)-(9)  which resulted in 1262 candidates. As we discussed in \textsection 2, the ALLWISE data base has a high completeness for finding quasars with $z-$band magnitudes brighter than 19.5. Limiting candidates to $z\le 19.5$ reduces our candidate list to  420 objects. Before conducting spectroscopic observations we visually inspected images of these 420 candidates and removed 231 candidates with suspicious detections, such as those close to very bright stars or binaries. This list of 189 objects is our primary $z\sim5$ quasar candidate sample. Removing 78 previously known quasars and one known dwarf results in a total of 110 candidates that required spectroscopic followup observations. We have obtained spectra for  99 of these candidates and also re-identified one known quasar (J022112.62$-$034252.26, See Table 3) that was not published at the time of observations. In addition to our primary sample at $z\le 19.5$, we also include candidates fainter than 19.5 as a supplementary sample, and observed several of them during our spectroscopic runs as a test for our ability to find fainter quasars using SDSS/{\it WISE} selection. 

\subsection{Spectroscopic Observations}

\begin{deluxetable*}{ccrrrrrrrrcrl}
\tabletypesize{\scriptsize}
\tablecaption{Observational Information of 72 newly identified $z\sim5$ quasars. \label{tbl-3}}
\tablewidth{0pt}
\tablehead{
\colhead{Name} &\colhead{Telescope} & \colhead{Grating} & \colhead{Slit} &
\colhead{Redshift} & \colhead{Date} & 
\colhead{Exposure (s)}
}

\startdata
  J000754.08$-$031730.82 & ANU & R3000 & 1.0 &4.76 & 20141016 & 1800\\
  J000851.43$$+$$361613.49 & LJT & G12 & 1.8 &5.17 & 20131127 & 2100\\
  J002526.84$-$014532.51 & MMT & 270GPM & 1.0 &5.07 & 20140110 & 600\\
  J003941.03$+$202554.85 & 216 & G4 & 2.3 &4.61 & 20141117 & 3600\\
  J004508.81$+$374334.91 & LJT & G5 & 1.8 &4.62 & 20141003 & 1800\\
  J005527.18$+$122840.67 & LJT & G3 & 1.8 &4.70 & 20131125 & 1800\\
  J011614.30$+$053817.70 & Bok & R400 & 2.5 &5.33 & 20141028 & 2100\\
  J012026.86$+$223058.55 & MMT & 270GPM & 1.5 &4.59 & 20140110 & 600\\
  J012220.29$+$345658.43 & LJT & G5 & 1.8 &4.85 & 20140121 & 1800\\
  J012247.35$+$121624.06 & LJT & G5 & 1.8 &4.79 & 20141024 & 2400\\
  J013127.34$-$032100.19 & LJT & G3 & 1.8 &5.18 & 20131125 & 1500\\
  J013224.89$-$030718.45 & ANU & R3000 & 1.0 &4.83 & 20150720 & 1200\\
  J013238.33$+$292602.57 & Bok & R400 & 2.5 &4.45 & 20141030 & 2400\\
  J014741.53$-$030247.88 & 216 & G4 & 2.3 &4.75 & 20141117 & 1800\\
  J015533.28$+$041506.74 & Bok & R400 & 2.5 &5.37 & 20141028 & 2400\\
  J015618.99$-$044139.88 & MMT & 270GPM & 1.5 &4.94 & 20140110 & 600\\
  J020139.04$+$032204.73 & 216 & G4 & 2.3 &4.57 & 20141117 & 3600\\
  J021624.16$+$230409.47 & Bok & R400 & 2.5 &5.26 & 20141030 & 3000\\
  J021736.76$+$470826.48 & MMT & 270GPM & 1.0 &4.81 & 20140108 & 600\\
  J022055.59$+$473319.34 & 216 & G4 & 2.3 &4.82 & 20141116 & 3600\\
  J022112.62$-$034252.26 & LJT & G3 & 1.8 &5.02 & 20131125 & 2700\\
  J024601.95$+$035054.12 & LJT & G5 & 1.8 &4.96 & 20140121 & 1800\\
  J024643.78$+$061045.74 & Bok & R400 & 2.5 &4.57 & 20141028 & 2100\\
  J025121.33$+$033317.42 & LJT & G3 & 1.8 &5.00 & 20131125 & 2400\\
  J030642.51$+$185315.85 & LJT & G3 & 1.8 &5.36 & 20131125 & 1320\\
  J032407.69$+$042613.29 & Bok & R400 & 2.5 &4.72 & 20141028 & 2100\\
  J045427.96$-$050049.38 & Bok & R400 & 2.5 &4.93 & 20141117 & 1500\\
  J065330.25$+$152604.71 & MMT & 270GPM & 1.0 &4.90 & 20140109 & 600\\
  J073231.28$+$325618.33 & MMT & 270GPM & 1.0 &4.76 & 20150509 & 300\\
  J074749.18$+$115352.46 & LJT & G3 & 1.8 &5.26 & 20131127 & 2100\\
  J075332.01$+$101511.68 & Bok & R400 & 2.5 &4.89 & 20141111 & 2400\\
  J080306.19$+$403958.96 & Bok & R400 & 2.5 &4.79 & 20141118 & 2400\\
  J083832.31$-$044017.47 & LJT & G12 & 1.8 &4.75 & 20140225 & 1800\\
  J085942.62$+$443115.97 & MMT & 270GPM & 1.0 &4.57 & 20150314 & 200\\
  J111700.43$-$111930.63 & LJT & G5 & 1.8 &4.40 & 20150221 & 1800\\
  J120829.27$+$394339.72 & MMT & 270GPM & 1.0 &4.94 & 20120527 & 300\\
  J122342.16$+$183955.39 & MMT & 270GPM & 1.0 &4.55 & 20150508 & 300\\
  J133257.45$+$220835.91 & MMT & 270GPM & 1.0 &5.11 & 20140109 & 600\\
  J143704.81$+$070807.71 & LJT & G5 & 1.8 &4.93 & 20150214 & 1800\\
  J152302.90$+$591633.04 & 216/MMT & 270GPM & 1.0 &5.11 & 20150314 & 600\\
  J155657.36$-$172107.55 & LJT & G5 & 1.8 &4.75 & 20150228 & 1500\\
  J160111.16$-$182835.08 & MMT & 270GPM & 1.0 &5.06 & 20150508 & 300\\
  J162045.64$+$520246.65 & MMT & 270GPM & 1.5 &4.79 & 20130418 & 900\\
  J162315.28$+$470559.90 & LJT & G5 & 1.8 &5.13 & 20140405 & 2100\\
  J162838.83$+$063859.14 & MMT & 270GPM & 1.0 &4.85 & 20150313 & 500\\
  J163810.39$+$150058.26 & MMT & 270GPM & 1.0 &4.76 & 20140514 & 900\\
  J165635.46$+$454113.55 & LJT & G5 & 1.8 &5.34 & 20141001 & 1800\\
  J175114.57$+$595941.47 & 216 & G10 & 2.3 &4.83 & 20140507 & 3600\\
  J175244.10$+$503633.05 & MMT & 270GPM & 1.5 &5.02 & 20130418 & 900\\
  J205442.21$+$022952.02 & MMT & 270GPM & 1.0 &4.56 & 20150508 & 300\\
  J211105.62$-$015604.14 & LJT & G5 & 1.8 &4.85 & 20140708 & 1547\\
  J215216.10$+$104052.44 & LJT & G5 & 1.8 &4.79 & 20141001 & 1200\\
  J220106.63$+$030207.71 & LJT & G5 & 1.8 &5.06 & 20141001 & 1500\\
  J220226.77$+$150952.38 & 216 & G4 & 2.3 &5.07 & 20141117 & 3000\\
  J220710.12$-$041656.28 & LJT & G5 & 1.8 &5.53 & 20141022 & 2400\\
  J221232.06$+$021200.09 & LJT & G5 & 1.8 &4.61 & 20141023 & 2400\\
  J221921.74$+$144126.31 & LJT & G5 & 1.8 &4.59 & 20141025 & 2400\\
  J222514.38$+$033012.50 & Bok & R400 & 2.5 &5.24 & 20141029 & 2400\\
  J222612.41$-$061807.29 & LJT & G5 & 1.8 &5.08 & 20141001 & 1500\\
  J225257.46$+$204625.22 & LJT & G5 & 1.8 &4.91 & 20141001 & 1500\\
  J232939.30$+$300350.78 & LJT & G5 & 1.8 &5.24 & 20141022 & 2100\\
  J233048.79$+$292301.05 & LJT & G5 & 1.8 &4.79 & 20141023 & 2400\\
  J234241.13$+$434047.46 & LJT & G5 & 1.8 &4.99 & 20141025 & 2400\\
  J234433.50$+$165316.48 & LJT & G5 & 1.8 &5.00 & 20140930 & 1500\\
 \hline
  J003125.86$+$071036.92 & LJT & G3 & 1.8 &5.33 & 20131126 & 3000\\
  J011546.27$-$025312.24 & LJT & G5 & 1.8 &5.07 & 20141123 & 3600\\
  J024152.92$+$043553.46 & LJT & G12 & 1.8 &5.22 & 20131129 & 2100\\
  J081248.82$+$044056.54 & Bok & R400 & 2.5 &5.29 & 20141030 & 3000\\
  J132319.69$+$291755.75  & LJT & G12 & 1.8 &4.92 & 20140223 & 2100\\
  J151901.27$+$042348.60 & MMT & 270GPM & 1.0 &4.94 & 20150316 & 600\\
  J165951.03$+$323928.63 & LJT & G12 & 1.8 &5.17 & 20140226 & 2400\\
  J215904.97$+$050745.76 & LJT & G5 & 1.8 &4.71 & 20141017 & 1800
\enddata

\tablecomments{The sources in the first part are from our main sample and those in the second part are from our fainter supplementary sample.}
\end{deluxetable*}

\begin{figure*}
\includegraphics[width=1.0\textwidth]{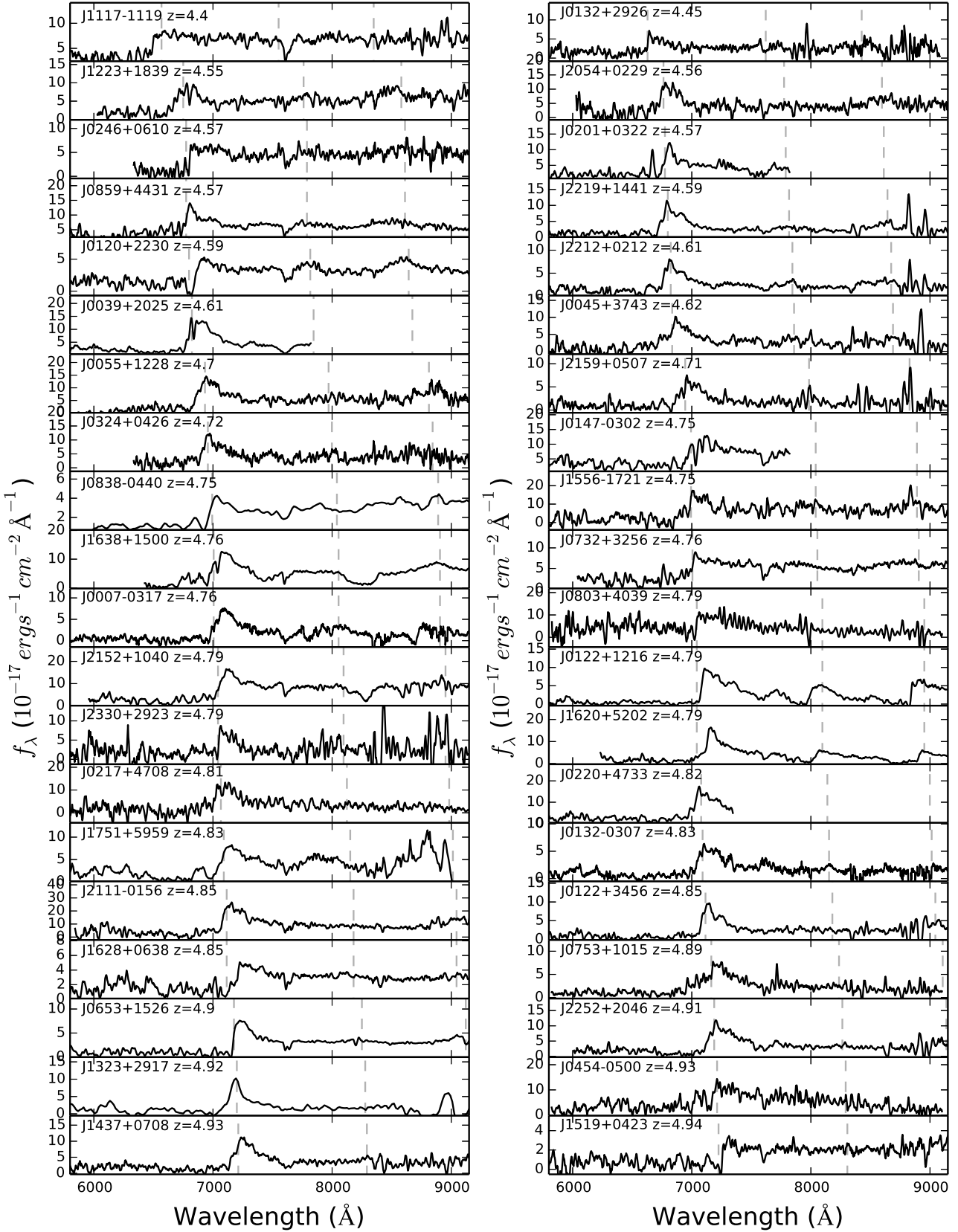}
\caption{Spectra of 72 newly discovered quasars. We smoothed the spectra by different size boxcars for spectra taken by different instruments. The spectra taken by WiFeS were smoothed to $R\sim200$. The spectra taken by G3 were smoothed to $R\sim100$. The spectra taken by G5 were smoothed to $R\sim110$. The MMT/270GPM spectra were smoothed to $R\sim130$ and $R\sim90$ for those using the $1\farcs0$ and $1\farcs5$ slits, respectively. The spectra taken by R400 were smoothed to $R\sim90$ and the spectra taken by G12 were smoothed to$R\sim50$. We present the spectra with the nominal flux calibrations obtained from standard star observations and scaled to the SDSS $i$-band magnitude. The vertical gray lines mark the locations of typical emission lines; in order,  $\rm Ly \alpha$, Si\,{\sc iv} and C\,{\sc iv}.
\label{fig6}}
\end{figure*}

\begin{figure*}
\figurenum{6}
\includegraphics[width=0.95\textwidth]{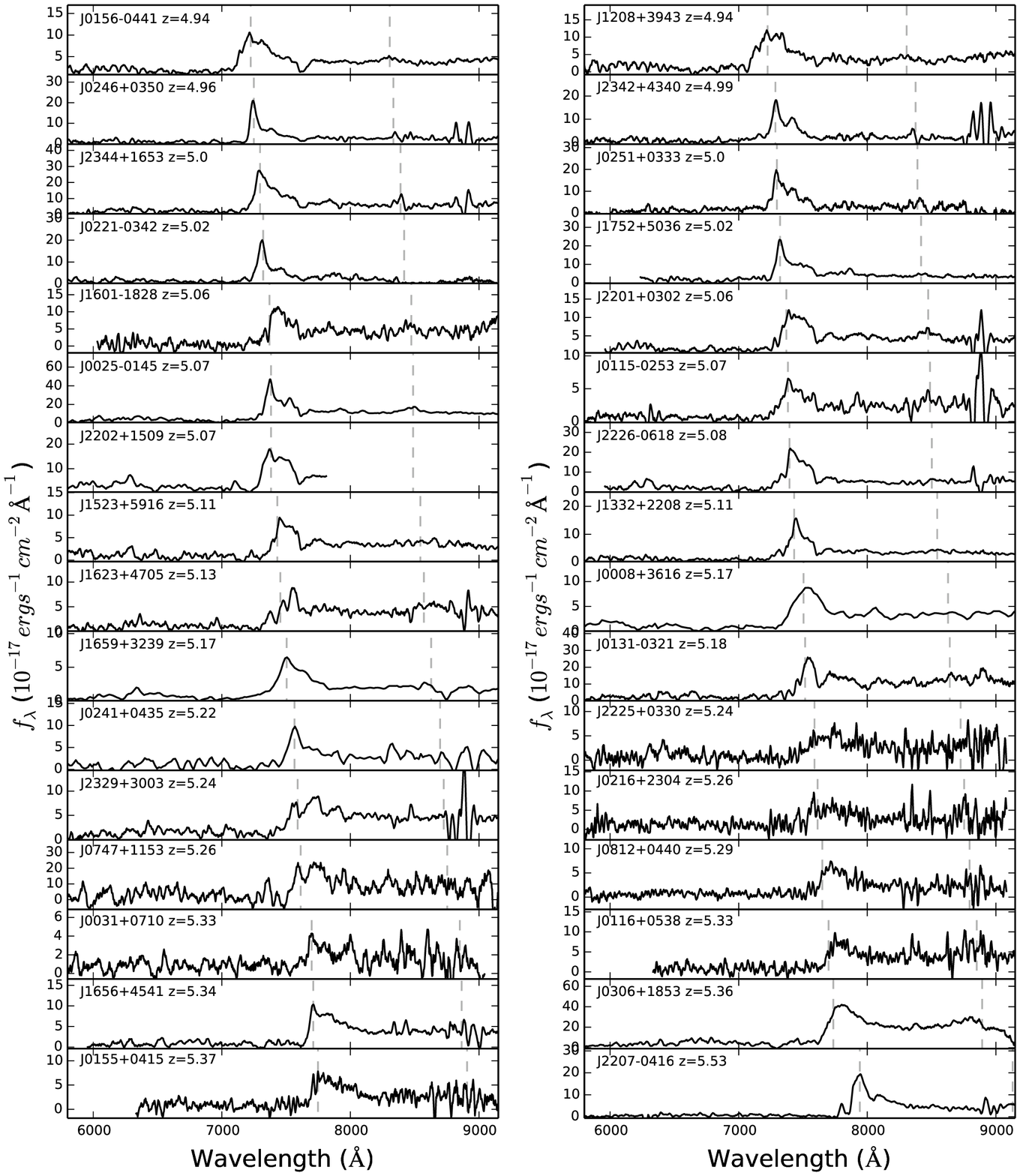}
\caption{Continued. \label{fig6b}}
\end{figure*}

Optical spectroscopic observations  to identify these quasar candidates were carried out using several facilities: the Lijiang 2.4 m telescope (LJT) and the Xinglong 2.16m telescope in China; the Kitt Peak 2.3m Bok telescope and the 6.5m MMT telescope in the U.S.; and the 2.3m ANU telescope in Australia. We have observed  99  candidates from our main sample and 64 (64.6\%) of them are high-redshift quasars with a redshift of $4.4 \lesssim z \lesssim 5.5$. We also observed 14 fainter candidates from our fainter candidates sample and 8 (57.1\%) of them are quasars at $4.7<z<5.4$. One of our candidates (J135457.62$+$314851.4) in our bright main sample was identified to be a low redshift low-ionization broad absorption line quasar (FeLoBAL QSO). However, we can not give the accurate redshift due to strong iron absorptions. The other 40 spectroscopic observed candidates are not quasars and were either identified as cool dwarfs or had relative low S/N and could only be ruled out as quasars.
Table 2 lists the observational information of the 72 new identified quasars.

The Lijiang 2.4m telescope is located at Lijiang Observatory, Yunnan Observatories, Chinese Academy of Sciences (CAS). It is equipped with the Yunnan Faint Object Spectrograph and Camera (YFOSC) which can take spectra followed by photometric images with a very short switching time. We observed 48 candidates by using the YFOSC, with a 2k $\times$ 4k CCD detector and three different grisms based on the brightness of our candidates. We used Grism 3 (G3) with dispersion of 172 $\rm \AA$/mm and wavelength coverage from 3200 to 9200 $\rm \AA$ to observe the brightest candidates; Grism 5 (G5), with dispersion of 185 $\rm \AA$/mm and wavelength coverage from 5000 to 9800 $\rm \AA$, to observe fainter candidates; and Grism 12 (G12), with dispersion of 900 $\rm \AA$/mm and wavelength coverage from 5600 to 9900 $\rm \AA$ to observe the faintest candidates in our sample. We used a $1\farcs8$ slit for all three grisms. This slit yields a resolution of $R\sim670$, $R\sim550$ and $R\sim160$ for the G3, G5 and G12 grisms, respectively. 

We observed 35 candidates using the Red Channel spectrograph \citep{schmidt89} on the MMT 6.5 m telescope. We used the 270$\rm l  mm^{-1}$ grating centered at 7500 $\rm \AA$, providing coverage from 5500 to 9700 $\rm \AA$. We used  the $1\farcs0$ or $1\farcs5$ slits based on the seeing, providing resolutions of $R\sim 640$ and $R\sim 430$, respectively. 

We observed 16 candidates using the Boller and Chivens Spectrograph (B\&C) on Steward Observatory's 2.3m Bok Telescope at Kitt Peak with the G400 Grating and $2\farcs5$ slit which gave a resolution of $R\sim 450$ and $\sim3400 \AA$ wavelength coverage. 

We observed eight candidates using the BAO Faint Object Spectrograph and Camera (BFOSC) on the 2.16 m optical telescope at the Xinglong station of the National Astronomical Observatories, Chinese Academy of Sciences (NAOC). We used the G4 or G10 gratings with dispersion of 198 $\rm \AA$/mm and 392 $\rm \AA$/mm, respectively. The wavelength coverage of these two gratings are 4000-7800 $\rm \AA$ and 4300-9000 $\rm \AA$ with spectral resolutions of $R\sim330$ and $R\sim110$ with a $2\farcs3$ slit, respectively. Note that we also one of the candidates (J1523+5916) using the MMT.

We also used the Wide Field Spectrograph \citep[WiFeS;][]{dopita07,dopita10}, an integral-field double-beam image-slicing spectrograph on the ANU 2.3m Telescope at Siding Spring Observatory, to observe seven of our quasar candidates. They were observed using Grating R3000 on WiFeS which gives a resolution of $R=3000$ at wavelengths‎ between 5300$\rm \AA$ and 9800 $\rm \AA$. 

All spectra taken by the 2.4m telescope, 2.16m telescope, 2.3m Bok telescope, and MMT telescope were reduced using standard IRAF routines. The WiFeS data were reduced with a python based pipeline PyWiFeS \citep{childress14}. The flux calibrations of all spectra were obtained from standard star observations on the same night and scaled to SDSS $i$-band magnitudes for absolute flux calibrations.

\section{Results}

\begin{deluxetable*}{ccrrrrrrrrrrrrrrcrl}
\tabletypesize{\scriptsize}
\tablecaption{Photometric Properties of 72 new identified $z\sim5$ quasars. \label{tbl-4}}
\tablewidth{0pt}
\tablehead{
\colhead{Name} & \colhead{Redshift} & \colhead{$m_{1450}$} & \colhead{$M_{1450}$} &
\colhead{$r$} &\colhead{$\sigma_{r}$}&\colhead{$i$} &\colhead{$\sigma_{i}$}& \colhead{$z$} &\colhead{$\sigma_{z}$}& 
\colhead{$W1$} &\colhead{$\sigma_{W1}$}& \colhead{$W2$} &\colhead{$\sigma_{W2}$}
}
\startdata
J000754.08$-$031730.82&4.76 &19.77 &-26.54 &21.64 &0.08 &19.63 &0.03 &19.49 &0.07 &16.26 &0.07 &15.41 &0.11 \\
J000851.43$+$361613.49&5.17 &19.12 &-27.34 &21.45 &0.08 &19.50 &0.02 &19.20 &0.05 &16.05 &0.05 &15.37 &0.09 \\
J002526.84$-$014532.51&5.07 &17.79 &-28.63 &19.58 &0.02 &18.03 &0.02 &17.85 &0.02 &14.80 &0.03 &14.16 &0.05 \\
J003941.03$+$202554.85&4.61 &19.26 &-27.00 &20.51 &0.05 &18.96 &0.03 &18.78 &0.05 &14.88 &0.03 &14.36 &0.05 \\
J004508.81$+$374334.91&4.62 &19.40 &-26.87 &20.43 &0.04 &19.37 &0.03 &19.06 &0.08 &15.98 &0.05 &15.31 &0.08 \\
J005527.18$+$122840.67&4.70 &18.85 &-27.45 &20.23 &0.03 &18.71 &0.02 &18.66 &0.04 &15.45 &0.05 &14.95 &0.09 \\
J011614.30$+$053817.70&5.33 &18.84 &-27.66 &21.57 &0.09 &19.87 &0.03 &19.22 &0.06 &16.37 &0.07 &15.76 &0.13 \\
J012026.86$+$223058.55&4.59 &19.38 &-26.88 &20.58 &0.04 &19.38 &0.02 &19.28 &0.06 &16.73 &0.09 &15.91 &0.17 \\
J012220.29$+$345658.43&4.85 &19.63 &-26.72 &21.30 &0.07 &19.45 &0.03 &19.45 &0.07 &16.52 &0.07 &15.69 &0.12 \\
J012247.35$+$121624.06&4.79 &19.54 &-26.79 &22.25 &0.14 &19.37 &0.03 &19.27 &0.06 &15.59 &0.05 &14.91 &0.07 \\
J013127.34$-$032100.19&5.18 &18.09 &-28.37 &20.15 &0.04 &18.46 &0.02 &18.01 &0.03 &14.58 &0.03 &13.84 &0.04 \\
J013224.89$-$030718.45&4.83 &19.74 &-26.60 &21.36 &0.06 &19.73 &0.03 &19.49 &0.06 &16.72 &0.09 &16.12 &0.17 \\
J013238.33$+$292602.57&4.45 &19.64 &-26.57 &20.73 &0.06 &19.64 &0.03 &19.47 &0.08 &16.50 &0.07 &15.82 &0.13 \\
J014741.53$-$030247.88&4.75 &18.55 &-27.77 &20.08 &0.03 &18.53 &0.02 &18.21 &0.02 &14.86 &0.03 &14.32 &0.05 \\
J015533.28$+$041506.74&5.37 &19.48 &-27.03 &21.70 &0.10 &19.97 &0.03 &19.26 &0.06 &16.33 &0.07 &15.19 &0.10 \\
J015618.99$-$044139.88&4.94 &19.21 &-27.17 &20.77 &0.04 &19.10 &0.02 &19.13 &0.05 &15.36 &0.04 &14.69 &0.06 \\
J020139.04$+$032204.73&4.57 &19.15 &-27.10 &20.25 &0.03 &19.09 &0.02 &19.02 &0.04 &15.38 &0.04 &14.83 &0.07 \\
J021624.16$+$230409.47&5.26 &19.30 &-27.18 &21.26 &0.06 &19.78 &0.03 &19.32 &0.06 &16.56 &0.08 &15.73 &0.15 \\
J021736.76$+$470826.48&4.81 &19.31 &-27.03 &20.55 &0.05 &18.96 &0.02 &18.88 &0.05 &15.76 &0.05 &15.14 &0.08 \\
J022055.59$+$473319.34&4.82 &18.56 &-27.78 &20.07 &0.03 &18.34 &0.01 &18.31 &0.03 &15.19 &0.04 &14.62 &0.06 \\
J022112.62$-$034252.26\tablenotemark{a}&5.02 &19.96 &-26.45 &20.86 &0.05 &19.25 &0.04 &19.50 &0.07 &16.38 &0.06 &15.63 &0.11 \\
J024601.95$+$035054.12&4.96 &19.46 &-26.93 &21.05 &0.05 &19.28 &0.02 &19.36 &0.05 &16.67 &0.07 &15.74 &0.14 \\
J024643.78$+$061045.74&4.57 &19.06 &-27.19 &20.22 &0.03 &19.05 &0.02 &18.86 &0.06 &15.42 &0.04 &14.81 &0.07 \\
J025121.33$+$033317.42&5.00 &19.58 &-26.82 &20.80 &0.04 &19.04 &0.03 &19.06 &0.05 &15.64 &0.04 &14.93 &0.07 \\
J030642.51$+$185315.85&5.36 &17.59 &-28.92 &19.89 &0.03 &17.96 &0.01 &17.47 &0.02 &14.31 &0.03 &13.46 &0.04 \\
J032407.69$+$042613.29&4.72 &19.19 &-27.12 &20.39 &0.04 &19.03 &0.02 &19.15 &0.06 &15.72 &0.05 &15.13 &0.09 \\
J045427.96$-$050049.38&4.93 &18.84 &-27.54 &19.91 &0.03 &18.59 &0.03 &18.39 &0.03 &15.09 &0.03 &14.53 &0.05 \\
J065330.25$+$152604.71&4.90 &19.35 &-27.02 &21.27 &0.06 &19.48 &0.02 &19.39 &0.07 &16.65 &0.11 &15.79 &0.16 \\
J073231.28$+$325618.33&4.76 &18.78 &-27.53 &20.26 &0.03 &18.82 &0.01 &18.62 &0.03 &15.46 &0.04 &14.92 &0.08 \\
J074749.18$+$115352.46&5.26 &18.51 &-27.97 &20.44 &0.03 &18.67 &0.02 &18.27 &0.03 &14.64 &0.03 &13.79 &0.04 \\
J075332.01$+$101511.68&4.89 &19.79 &-26.57 &21.14 &0.04 &19.39 &0.02 &19.37 &0.06 &16.31 &0.08 &15.78 &0.15 \\
J080306.19$+$403958.96&4.79 &19.17 &-27.15 &20.58 &0.04 &18.88 &0.02 &18.60 &0.03 &15.28 &0.04 &14.76 &0.06 \\
J083832.31$-$044017.47&4.75 &19.37 &-26.94 &21.20 &0.06 &19.62 &0.03 &19.21 &0.07 &15.58 &0.04 &15.06 &0.08 \\
J085942.62$+$443115.97&4.57 &18.66 &-27.59 &19.67 &0.02 &18.66 &0.03 &18.64 &0.04 &15.60 &0.05 &15.06 &0.08 \\
J111700.43$-$111930.63&4.40 &18.61 &-27.58 &19.75 &0.02 &18.65 &0.02 &18.29 &0.03 &15.19 &0.04 &14.63 &0.06 \\
J120829.27$+$394339.72&4.94 &19.20 &-27.18 &20.79 &0.06 &19.04 &0.02 &19.06 &0.05 &15.80 &0.05 &15.09 &0.08 \\
J122342.16$+$183955.39&4.55 &18.90 &-27.34 &20.52 &0.04 &18.97 &0.02 &18.59 &0.04 &14.98 &0.03 &14.45 &0.05 \\
J133257.45$+$220835.91&5.11 &19.11 &-27.32 &21.12 &0.04 &19.26 &0.02 &19.23 &0.04 &15.69 &0.05 &14.89 &0.06 \\
J143704.81$+$070807.71&4.93 &19.35 &-27.03 &20.62 &0.04 &19.17 &0.02 &19.16 &0.05 &16.14 &0.06 &15.62 &0.12 \\
J152302.90$+$591633.04&5.11 &19.10 &-27.33 &21.39 &0.06 &19.54 &0.02 &19.22 &0.05 &15.64 &0.03 &15.13 &0.05 \\
J155657.36$-$172107.55&4.75 &18.47 &-27.85 &19.94 &0.04 &18.43 &0.02 &18.43 &0.05 &15.09 &0.04 &14.59 &0.06 \\
J160111.16$-$182835.08&5.06 &18.96 &-27.46 &20.98 &0.15 &19.37 &0.05 &18.89 &0.09 &15.65 &0.05 &15.05 &0.08 \\
J162045.64$+$520246.65&4.79 &19.09 &-27.24 &20.77 &0.04 &18.97 &0.02 &18.94 &0.04 &15.30 &0.03 &14.70 &0.04 \\
J162315.28$+$470559.90&5.13 &18.89 &-27.55 &20.87 &0.05 &19.52 &0.03 &19.23 &0.07 &15.57 &0.03 &14.76 &0.05 \\
J162838.83$+$063859.14&4.85 &19.37 &-26.98 &20.88 &0.04 &19.56 &0.02 &19.40 &0.05 &16.68 &0.09 &15.93 &0.17 \\
J163810.39$+$150058.26&4.76 &18.81 &-27.50 &20.53 &0.04 &18.83 &0.02 &18.53 &0.04 &15.10 &0.04 &14.53 &0.05 \\
J165635.46$+$454113.55&5.34 &18.94 &-27.57 &21.51 &0.06 &19.70 &0.02 &19.06 &0.04 &16.22 &0.28 &15.53 &0.07 \\
J175114.57$+$595941.47&4.83 &19.03 &-27.17 &20.75 &0.04 &19.09 &0.02 &18.78 &0.04 &15.66 &0.03 &15.09 &0.05 \\
J175244.10$+$503633.05&5.02 &18.97 &-27.43 &20.85 &0.04 &18.82 &0.02 &18.87 &0.05 &15.13 &0.03 &14.40 &0.03 \\
J205442.21$+$022952.02&4.56 &19.15 &-27.10 &20.33 &0.03 &19.20 &0.02 &19.00 &0.05 &16.33 &0.07 &15.78 &0.14 \\
J211105.62$-$015604.14&4.85 &18.21 &-28.14 &19.78 &0.02 &18.11 &0.02 &18.14 &0.03 &15.02 &0.04 &14.41 &0.05 \\
J215216.10$+$104052.44&4.79 &18.37 &-27.96 &19.97 &0.03 &18.36 &0.02 &18.22 &0.03 &14.67 &0.03 &14.02 &0.04 \\
J220106.63$+$030207.71&5.06 &18.90 &-27.52 &20.58 &0.03 &19.11 &0.02 &18.90 &0.04 &15.98 &0.06 &15.20 &0.10 \\
J220226.77$+$150952.38&5.07 &18.48 &-27.95 &20.28 &0.03 &18.69 &0.02 &18.47 &0.03 &15.74 &0.05 &15.20 &0.08 \\
J220710.12$-$041656.28&5.53 &18.86 &-27.70 &22.32 &0.24 &19.59 &0.03 &18.95 &0.06 &15.12 &0.04 &14.14 &0.05 \\
J221232.06$+$021200.09&4.61 &19.83 &-26.43 &20.90 &0.03 &19.68 &0.03 &19.41 &0.05 &16.65 &0.08 &15.86 &0.14 \\
J221921.74$+$144126.31&4.59 &19.69 &-26.57 &20.66 &0.05 &19.53 &0.03 &19.19 &0.06 &16.20 &0.08 &15.45 &0.12 \\
J222514.38$+$033012.50&5.24 &19.38 &-27.10 &21.74 &0.14 &20.02 &0.05 &19.47 &0.10 &16.50 &0.08 &15.69 &0.13 \\
J222612.41$-$061807.29&5.08 &18.66 &-27.76 &20.32 &0.04 &18.76 &0.02 &18.73 &0.05 &15.64 &0.05 &14.96 &0.09 \\
J225257.46$+$204625.22&4.91 &19.44 &-26.93 &20.65 &0.04 &19.16 &0.02 &19.23 &0.06 &16.27 &0.06 &15.52 &0.10 \\
J232939.30$+$300350.78&5.24 &18.83 &-27.65 &20.87 &0.05 &19.37 &0.02 &18.93 &0.04 &16.21 &0.06 &15.43 &0.10 \\
J233048.79$+$292301.05&4.79 &19.73 &-26.59 &20.93 &0.05 &19.53 &0.02 &19.37 &0.06 &16.77 &0.10 &15.80 &0.13 \\
J234241.13$+$434047.46&4.99 &19.54 &-26.86 &21.17 &0.06 &19.26 &0.02 &18.97 &0.05 &15.57 &0.04 &14.73 &0.06 \\
J234433.50$+$165316.48&5.00 &18.54 &-27.86 &20.23 &0.03 &18.46 &0.02 &18.52 &0.03 &15.22 &0.04 &14.56 &0.06 \\
\hline
J003125.86$+$071036.92&5.33 &20.21 &-26.29 &22.46 &0.15 &20.42 &0.04 &20.09 &0.09 &16.70 &0.10 &15.48 &0.12 \\
J011546.27$-$025312.24&5.07 &19.56 &-26.86 &21.28 &0.06 &19.88 &0.03 &19.58 &0.07 &16.42 &0.08 &15.87 &0.17 \\
J024152.92$+$043553.46&5.22 &19.40 &-27.07 &21.42 &0.08 &19.78 &0.03 &19.55 &0.08 &16.26 &0.06 &15.60 &0.13 \\
J081248.82$+$044056.54&5.29 &19.77 &-26.72 &21.85 &0.11 &20.05 &0.04 &19.77 &0.10 &16.31 &0.07 &15.55 &0.14 \\
J132319.69$+$291755.75&4.92 &20.16 &-26.22 &21.74 &0.10 &19.74 &0.04 &19.95 &0.11 &16.65 &0.11 &15.61 &0.15 \\
J151901.27$+$042348.60&4.94 &19.79 &-26.59 &21.63 &0.06 &20.22 &0.03 &19.83 &0.08 &16.37 &0.06 &15.59 &0.10 \\
J165951.03$+$323928.63&5.17 &19.80 &-26.65 &21.82 &0.07 &20.00 &0.02 &19.89 &0.07 &16.42 &0.06 &15.66 &0.10 \\
J215904.97$+$050745.76&4.71 &20.24 &-26.06 &21.14 &0.06 &19.74 &0.03 &19.55 &0.08 &16.81 &0.10 &15.95 &0.17 
\enddata
\tablenotetext{a}{This quasar was also independently discovered by SDSS DR12 (I. P\^{a}ris et al. (2016), in preparation).} 
\tablecomments{The sources in the first part are from our main sample and those in the second part are from our fainter supplementary sample.}
\end{deluxetable*}

\subsection{Discovery of 72 New Quasars at $z\sim$5}
We have spectroscopically observed 99 candidates with $z$-band magnitudes brighter than 19.5, and 64 (64.6\%) of them are quasars with redshifts of $4.4\lesssim z \lesssim 5.5$ and absolute magnitudes of $-29.0\lesssim M_{1450} \lesssim -26.4$. We also observed 14 fainter candidates  selected with the same selection criteria and identified 8 (57.1\%) fainter $z\sim5$ quasars with $4.7<z<5.4$ and absolute magnitude of $-27.1\lesssim M_{1450} \lesssim -26.1$.
Table 4 lists the redshifts and SDSS $+$ {\it WISE} photometry of the 72 newly discovered quasars and Figure 5 shows the optical spectra of these new quasars. The redshifts of these quasars were measured from $\rm Ly \alpha$, N\,{\sc v}, O\,{\sc i}/Si\,{\sc ii}, C\,{\sc ii}, Si\,{\sc iv}, and C\,{\sc iv} emission lines (any available). We used a visual recognition assistant for quasar spectra software  \citep[ASERA;][]{yuan13}, which is an interactive semi-automated toolkit allowing the user to visualize observed spectra and measure the redshifts by fitting the observed spectra to the SDSS quasar template \citep{vanden01}, and interactively accessing related spectral line information. The redshift error measured based on this method mainly depends on the quality of the observed spectra and line properties. Due to the low resolution and strong absorptions  blueward of $\rm Ly \alpha$, the typical redshift error is about 0.05 for our newly discovered quasars. However, the redshift error could be up to 0.1 for objects with relatively low S/N spectra. Our quasar sample spans a redshift range of $4.40\le z \le 5.53$ and the redshift distribution of these newly identified quasars is shown in the lower panel of Figure 7. As we discussed in Section 2, there is a gap in the previously published quasar redshift distribution at $5.2 < z <5.7$ with only 33 published quasars at this redshift due to low identification efficiency;  17 of them were identified by the SDSS quasar survey. Among the 72 newly identified quasars, 12 of them are at $5.2 < z <5.7$, which represents an increase of $\sim$36\% in the number of known quasars in this difficult redshift range.

In Table 3, Columns $m_{1450}$ and $M_{1450}$ list the apparent and absolute AB magnitudes of the continuum at rest-frame 1450$\rm \AA$, respectively. They were calculated by fitting a power-law continuum $f_\nu \sim  \nu^{\alpha_{\nu}}$ to the spectrum of each quasar. As many spectra of our newly discovered quasars do not have enough continuum coverage to reliably measure the slopes of the continua, we assumed that the slope is consistent with the average quasar UV continuum slope $\alpha_{\nu}=-0.5$ \citep{vanden01}.
The power-law continuum was then normalized to match visually identified continuum windows that contain minimal contribution from quasar emission lines and from sky OH lines. 
Figure 7 shows the absolute magnitude at rest-frame 1450$\rm \AA$ and the redshift distribution of our 72 newly discovered quasars and the published SDSS $z\ge4.5$ quasars. The red stars are our newly discovered quasars and the blue crosses denote SDSS quasars. The red and blue dashed lines in Figure 7 denote the mean absolute magnitude of our new quasars ($\overline {M}_{1450}=-27.2$) and the SDSS quasars ($\overline {M}_{1450}=-26.4$), respectively. Our newly discovered quasars are systematically brighter than SDSS quasars and improved the completeness of luminous $z\sim5$ quasars in the SDSS footprint. More importantly, 24 of our newly discovered quasars have $M_{1450}<-27.5$ and doubled the number of known quasars (26 $z>4.5$ SDSS quasars) in this brightness range in the SDSS footprint. In particular, 22 of our new quasars are at $z>4.7$ with $M_{1450}<-27.5$, compared with only 13 previously published SDSS quasars in this redshift/luminosity range.

\begin{figure}[tbh]
\includegraphics[width=0.5\textwidth]{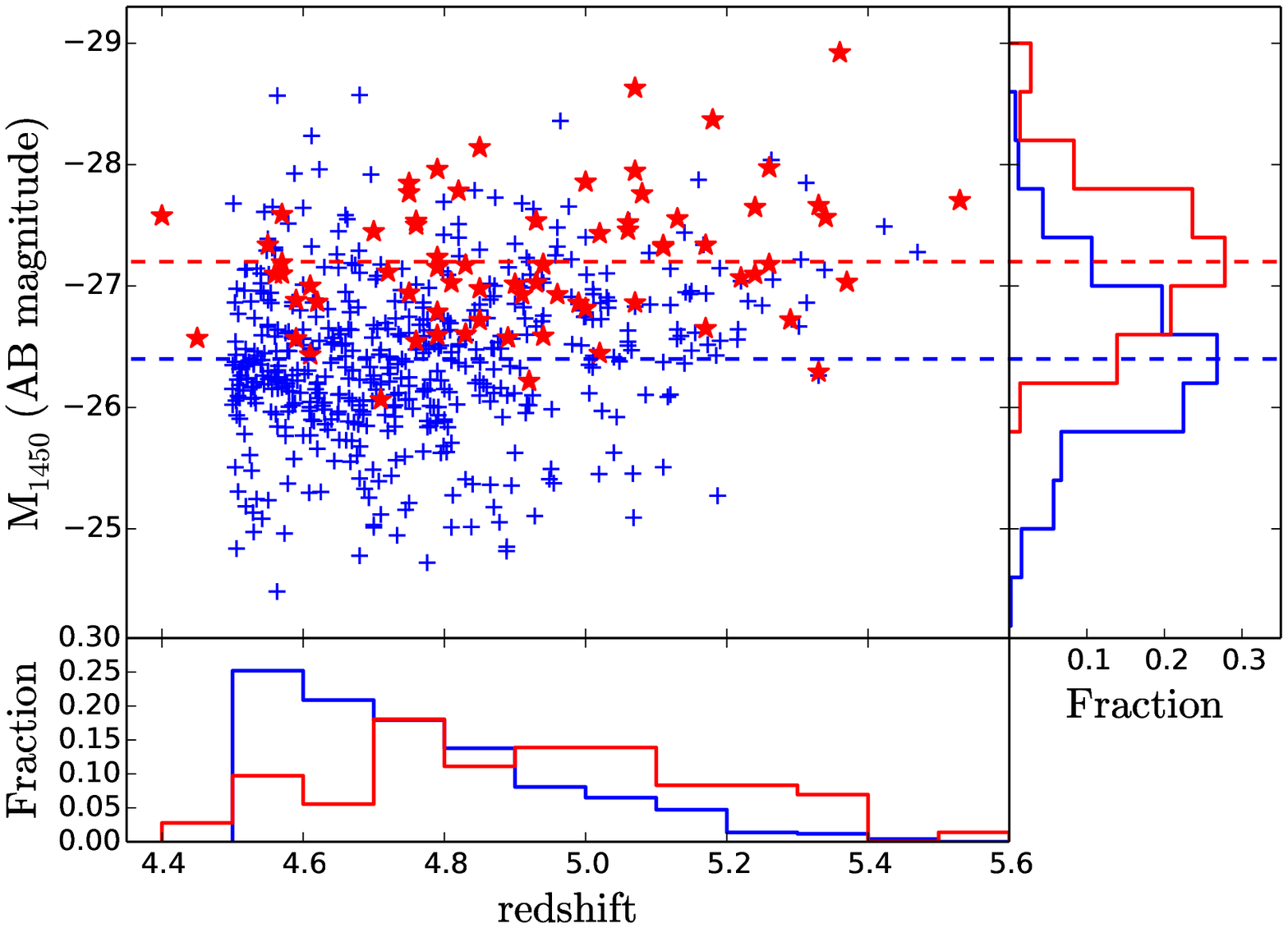}
\caption{Upper panel: The $\rm M_{1450}$ vs. redshift diagram. The small blue crosses denote SDSS $z\ge4.5$ quasars and the red stars denote our newly discovered quasars. The $\rm M_{1450}$ are AB magnitudes of quasars at 1450$\rm \AA$. Lower panel: The redshift distributions of newly identified quasars and known $z\ge4.5$ quasars. Our method improves the completeness of selecting quasars at $z>5.0$ which is consistent with the prediction from the color-$z$ relation shown in Figure 3. Right panel: The distribution of $\rm M_{1450}$. Apparently our newly discovered quasars are systematically brighter than SDSS quasars and improved the completeness of luminous $z\sim5$ quasars in the SDSS footprint.
\label{fig6}}
\end{figure}

\subsection{Notes on Individual Objects}

SDSS J013127.34$-$032100.1 ($z=5.19$). The radio-loudness defined as the ratio of the rest-frame flux densities in the radio (5GHz) to optical bands (4400$\rm \AA$) bands \citep{kellermann89}. J0131-0321 is a radio-loud quasar with radio loudness about 100. J0131$-$0321 is the most luminous $z\gtrsim5$ radio-loud quasar  known, with SDSS $z$-band magnitude $18.01\pm0.03$ and with $M_{1450}=-28.29$. The observational properties of this quasar are discussed in detail in a separate paper \citep{yi14}.

SDSS  J022112.62$-$034252.26 ($z=5.02$). J0221-0342 was independently discovered by the BOSS quasar survey and published in the DR12 quasar catalog (I. P{\^a}ris et al. 2016, in preparation).

SDSS J030642.51$+$185315.8 ($z=5.36$). J0306$+$1853 is the most luminous $z\gtrsim5$ quasar known to date,  with $M_{1450}=-28.92$. A more detailed analysis of this quasar is in \cite{wang15}.

SDSS J220710.12$-$041656.28 ($z=5.53$). J2207-0416 is the most distant quasar discovered in our $z\sim5$ main sample. Note that due to the extremely similar optical-to-IR colors of $z\sim5.5$ quasars and M dwarfs, there are only two known $z\sim5.5$ quasars published before: RD J030117$+$002025 at $z=5.50$ \citep{stern00} and NDWFS J142729.7$+$352209 at $z=5.53$ \citep{cool06}.

\section{Discussion}

\subsection{Comparison with SDSS $z\sim5$ Quasar Selection}
The SDSS quasar surveys provided the largest quasar sample selected based on SDSS $u,~g,~r,~i,~z$ photometry and have discovered $\sim500$ quasars at $z>4.5$ \citep[e.g.][]{schneider10,paris12,paris14}. The primary method for selecting $z>4.5$ quasars in the SDSS quasar surveys is based on the $r-i/i-z$ color$-$color diagram \citep{fan99,richards02}. 
Since the third stage of the SDSS (BOSS) high-redshift quasar survey mainly focused on fainter targets, here we  compare our selection only to the first two stages of SDSS high-redshift quasar selection. There are 392 $z>4.5$ quasars in the SDSS DR7 quasar catalog \citep{schneider10}. Three hundred and fifty-six of them have counterparts within $2"$ in the ALLWISE catalog and 126 (32\%) of them satisfy our selection criteria (also including those with $z$-band magnitudes fainter than 19.5). Figure 8 shows 392 SDSS high-$z$ quasars: red circles denote quasars that satisfy our selection criteria and blue crosses denote quasars that do not satisfy our selection criteria. Clearly quasars that are not selected by our method are mainly objects with $z\lesssim4.7$ quasars or with fainter magnitudes. As Figure 3 shows, about 60\% of quasars at $z\lesssim4.7$ have $W1-W2<0.5$, which is the reason why our method is not sensitive to $z\lesssim4.7$ quasars.

All our targets are within the SDSS footprint, therefore we examined the 72 newly discovered quasars against  the SDSS high-redshift quasar selection criteria presented in Richards et al. (2002). There are three reasons that they are not in the SDSS quasar catalogs:   (1) 38 of them are not in the SDSS DR7 photometry sky coverage and thus not in the SDSS main spectroscopy survey region; (2)  25 new quasars have the quasar target flag set by the latest SDSS target selection pipeline but were not observed either because these candidates are at the edge of SDSS main spectroscopy survey or the fields of these candidates were observed at the early stage of the SDSS (e.g. SDSS EDR and DR1) when a preliminary version of the selection pipeline was used (Richards et al. 2002), and (3) the other nine new quasars were not targeted by SDSS altogether.
These nine quasars were rejected by the photometric flags (e.g. BLENDED, INTERP\_CENTER, CHILD, and FAMILY).
Although the SDSS high-redshift quasar target selection method can recover many of these newly discovered quasars, the efficiency (less than 10\%) of the SDSS target selection is much lower than that ($\sim$60\%) of our method by combining SDSS and ALLWISE colors.  
In addition, our method improved both the completeness and efficiency for selecting $z\gtrsim 5.1$ quasars (the region between the purple dashed line and the orange dashed line in Figure 4).  

\begin{figure}
\includegraphics[width=0.5\textwidth]{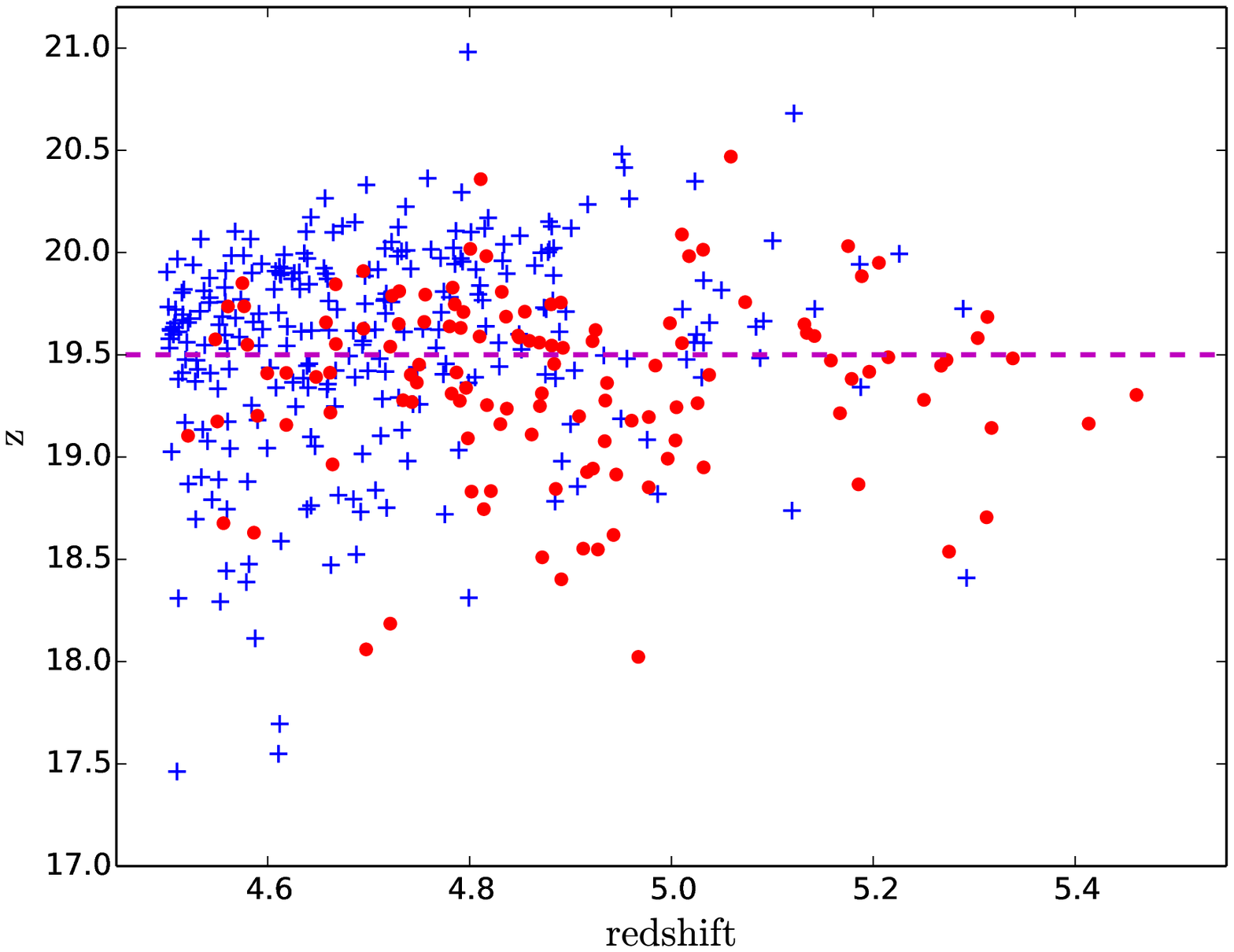}
\caption{Redshift vs. $z$-band magnitude diagram of SDSS DR7 $z>4.5$ quasars. The red circles denote quasars that satisfy our selection method while blue crosses denote quasars that do not satisfy our selection method. The purple line represents $z-$band magnitudes equal to 19.5.}
\label{fig8}
\end{figure}

\subsection{Efficiency and Completeness of Our Quasar Selection}
In our spectroscopic observations $\sim$60\% of our candidates are real high-redshift quasars, which is a relatively high efficiency for selecting high-redshift quasars. Although our optical color cuts are very sensitive to $4.5\lesssim z \lesssim 5.4$ (Figure 4), quasars at $z\lesssim4.7$ have a bluer $W1-W2$ color and lead to a lower completeness of our selection method in this redshift range (Figures 3 and Figure 8). 
At $z\sim5$, the {\it WISE} $W1$ and $W2$ bands only cover quasar rest-frame optical emission (shorter than $1\mu$m), which is not significantly polluted by the hot dust \citep[e.g.][]{leipski14}. The $W1-W2$ color changes a lot from redshift 4.5 to 5.5 as the $\rm H \alpha$ emission line moves from $W1$ to $W2$. More than half of quasars at $z\lesssim4.7$ have a bluer $W1-W2$ with $W1-W2<0.5$. At this redshift range, the $\rm H \alpha$ emission line redshifts into the $W1$ band. Quasars with weaker $\rm H \alpha$ emission lines or redder continuum will prefer redder $W1-W2$ colors and quasars with stronger $\rm H \alpha$ emission lines or bluer continuum will show bluer $W1-W2$ colors. So our $W1-W2$ cut will bias to quasars with weaker $\rm H \alpha$ emissions or redder continuum at $z\lesssim4.7$. At $z\gtrsim4.7$, the $\rm H \alpha$ emission line moves out from $W1$ band and the $W1-W2$ colors change to be redder and our $W1-W2$ cut will not bias to quasars emission line and continuum properties significantly.
Overall, our selection method can select luminous quasars at $4.7\lesssim z \lesssim 5.4$ with both high efficiency and relatively high completeness (e.g. Figure 7). Benefiting from this we are able to study the QLF at $z\sim5$, especially at the bright end of the QLF, and do detailed statistics on quasar properties based on a complete sample of SDSS$-${\it WISE} selected $z \sim 5$ quasars\citep{yang16}. More detailed analysis of the completeness of our quasar sample will be discussed in the QLF paper.

Although ALLWISE has whole-sky coverage, SDSS is limited to only one quarter of the sky.  New optical sky surveys are providing coverage of the entire high galactic latitude sky: the PAN-STARRS-1 survey \citep{kaiser02,kaiser10}  has a much larger sky coverage and a slightly deeper depth in red filters than SDSS and it has a $y$-band filter that covers the $\rm 9200\AA<\lambda<10500\AA$ range, allowing quasar selection to $z\sim 7$. Combining PAN-STARRS with ALLWISE photometry will provide a much more comprehensive way to find luminous high-redshift quasars. Many other ongoing large-area optical sky surveys including SkyMapper, VST ATLAS, DECaLS, BASS, and DES will enable a complete all-sky survey of  luminous high-redshift quasars by combining them with the ALLWISE photometry.

\section{Optical Surveys Plus {\it WISE} for Finding Luminous $z\gtrsim5.7$ Quasars}

\begin{deluxetable*}{ccrrrrrrrrrrl}
\tabletypesize{\scriptsize}
\tablecaption{Photometry of four $z>5.7$ quasars selected by our method. \label{tbl-5}}
\tablewidth{0pt}
\tablehead{
\colhead{Name} & \colhead{Redshift} &  \colhead{$m_{1450}$} & \colhead{$M_{1450}$} &
\colhead{$i$} &\colhead{$\sigma_{i}$}& \colhead{$z$} &\colhead{$\sigma_{z}$}& 
\colhead{$W1$} &\colhead{$\sigma_{W1}$}& \colhead{$W2$} &\colhead{$\sigma_{W2}$}
}
\startdata
J010013.02$+$280225.8\tablenotemark{a}& $6.30\pm0.01$ &17.51 &-29.26 &20.84 &0.06 &18.33 &0.03 &14.46 &0.03 &13.64 &0.03 \\
J154552.08$+$602824.0&$5.78 \pm0.03$ &19.26 &-27.37 &21.27 &0.07 &19.09 &0.05 &16.00 &0.04 &15.16 &0.05 \\
J232514.24$+$262847.6&$5.77 \pm0.05$ &19.64 &-26.98 &21.62 &0.17 &19.41 &0.10 &16.19 &0.06 &15.41 &0.10 \\
J235632.44$-$062259.2\tablenotemark{b}&$6.15 \pm0.02$ &19.89 &-26.85 &22.55 &0.35 &19.78 &0.11 &16.56 &0.10 &15.70 &0.20 
\enddata
\tablenotetext{a}{See \cite{wu15} for details.} 
\tablenotetext{b}{It was also independently discovered by the Pan-STARRS1 high redshift quasar survey (E. Ba\~nados et al. 2016, in preparation).}
\end{deluxetable*}

As Figure 3 shows, quasars at $4.7\lesssim z \lesssim 7$ have relatively redder $W1-W2$ colors which are useful for rejecting late-type star contamination. Figure 9 shows the known $z>5.0$ quasars (Table 1 and references therein) and M, L, and T dwarfs \citep{kirkpatrick11} on the $z-W1/W1-W2$ color$-$color diagram. The cyan cycles denote $5.0 < z < 5.5 $ quasars, the blue triangles represent $5.5<z<6.5$ quasars, and the orange stars represent $z>6.5$ quasars. Clearly the optical plus {\it WISE} photometric selection method can separate quasars from contaminants at redshifts up to $z\lesssim 6.4$, the upper redshift limit of optical surveys. 
However, the L and T dwarfs begin to overlap with $z\gtrsim6.5$ quasars in this color$-$color diagram which is a limit for finding quasars at this redshift range. It could be improved by combining near-infrared photometry (e.g. $Y$-band and $J$-band) as L and T dwarfs have redder $Y-J$ colors than that of $z\gtrsim6.5$ quasars \citep{venemans13}.
Actually, \cite{carnall15} have used a similar method ($z-W2/W1-W2$) to search for $z>5.7$ quasars in the VST ATLAS survey and have identified two new $z>6$ luminous quasars in the ATLAS survey area. 

\begin{figure}[tbh]
\includegraphics[width=0.5\textwidth]{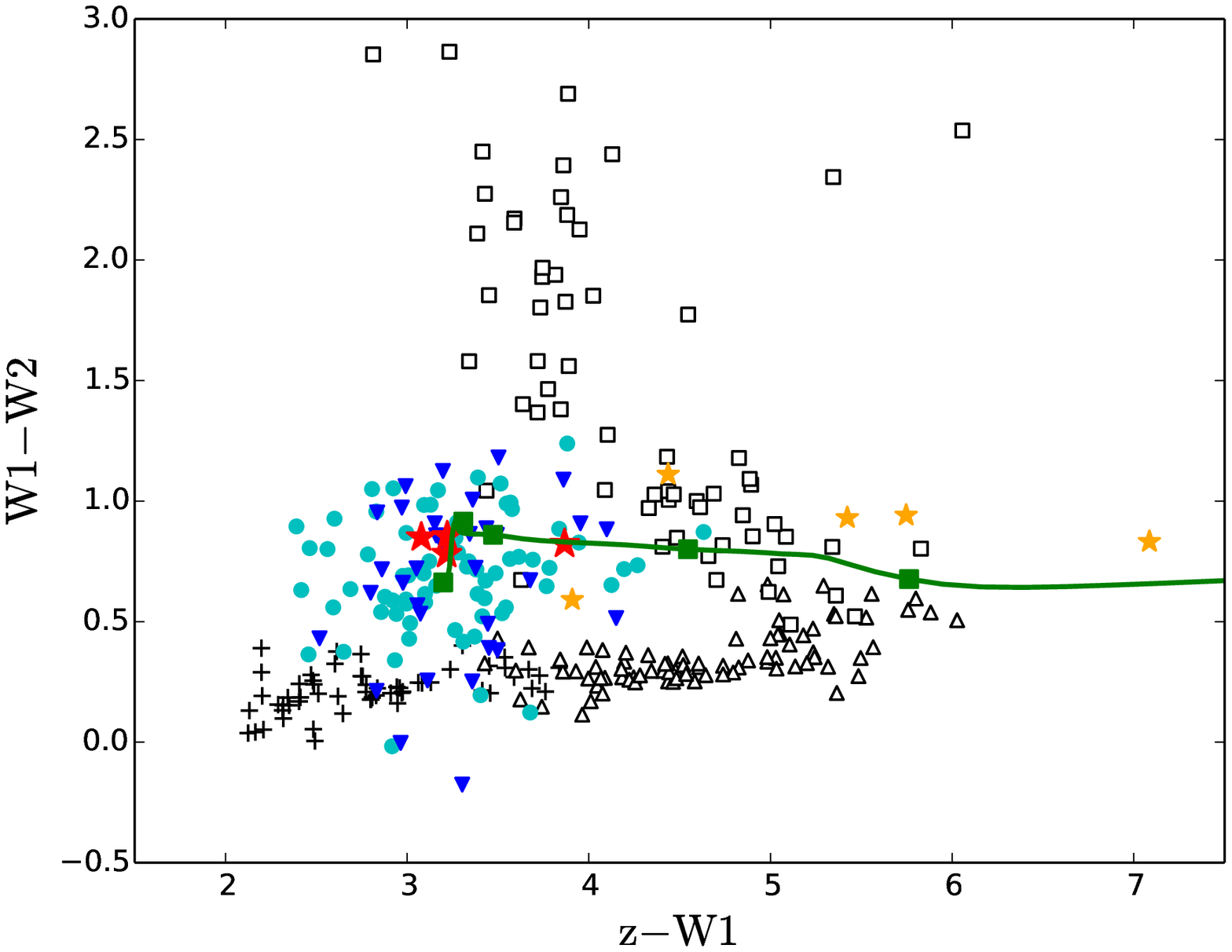}
\caption{The $z-W1$ vs. $W1-W2$ color$-$color diagram. The green solid line represents the color$-$z relation predicted using quasar composite spectra. The green solid squares mark the color tracks for quasars from $z=5.0$ to $z=7.0$, in steps of $\Delta z =0.5$.  The cyan circles denote $5\le z < 5.5$ quasars, the blue triangles denote $5.5 \le z <6.5$ quasars, and the orange stars denote $z\ge 6.5$ quasars. The red stars represent our four newly discovered  $z\gtrsim5.7$ quasars. The black crosses, open triangles, and open squares are M, L, and T dwarfs, respectively, from \cite{kirkpatrick11}. 
\label{fig9}}
\end{figure}

Motivated by the high detection rate of luminous high-redshift quasars as described in section 2 (e.g., $M_{1450}\lesssim-26.5$ at $z\lesssim6$ and $M_{1450}\lesssim-27.0$ at $z\lesssim7$), we also search for luminous $z\gtrsim5.7$ quasars by combining the $i$-dropout technique \citep[e.g.,][]{fan01b,willott07,jiang08,banados14} and $z-W1/W1-W2$ colors mentioned above. This method can be used to reject superposed objects without any photometric followup observations and can reject the majority of contaminants (mainly L and T dwarfs). We selected about 20 $i$-dropouts with $z_{AB}<19.8$ in the SDSS area. Before doing spectroscopy we visually compared the spectra energy distribution (SED) of these candidates with type 1 quasar composite SED \citep{richards06} to rank the priority.

We have observed four candidates with the highest priority and  have identified four $z\gtrsim5.7$ quasars: J010013.02$+$280225.8 at $z=6.30\pm0.01$ \citep{wu15}, J235632.44$-$062259.2 at $z=6.15\pm0.02$, J154552.08$+$602824.0 at $z=5.78\pm0.03$ and J232514.24$+$262847.6 at $z=5.77\pm0.05$. Table 4 lists the photometric and redshift information for these four quasars and Figure 10 shows their optical spectra. 
The spectrum of J0100+2802 was obtained with LBT/MODS spectrograph \citep{pogge10} and the redshift determined from Mg\,{\sc ii} emission line in the near-IR spectrum \citep{wu15}. J0100+2802 is the most luminous $z>6$ quasar ever known and hosts a 12 billion solar black hole in its center \citep{wu15}.
We observed J1545+6028 with the G5 grating on the LJT 2.4m telescope with a 3600-s exposure on 2014 April 5. This quasar was not targeted by SDSS main $z\sim6$ quasar survey as its $i-z=2.18$ does not satisfy their $i-z$ cut \citep{fan01a}.
The spectrum of J2325+2628 was obtained with MMT Red Channel on 2015 May 9. We used the 270GPM grating centered at 8500$\rm \AA$ and a $1\farcs0$ slit. The MMT spectrum shows that J2325+2628 is a weak-line emission quasar (WLQ) and with the current S/N there are no clear emission lines. The redshift of J2325+2628 was estimated by fitting the continuum to the composite spectra \citep{vanden01}. Although this quasar has SDSS photometry data and satisfy SDSS $z\gtrsim5.7$ selection criteria, it does not fall in the SDSS $z\gtrsim5.7$ quasar survey region.
We observed J2356$-$0622 with ANU WiFeS with a 30-minute exposure on 2015 May 15. However, the target was only marginally detected due to the cloudy weather and it was hard to confirm that it is a high-redshift quasar. A further one-hour exposure was obtained with WiFeS on 20 July, which confirmed J2356$-$0622 as a quasar at $z=6.15$. J2356$-$0622 is the faintest $z>5.7$ quasar in our sample with $M_{1450}=-26.85$. This quasar has a bad SDSS photometry flag (maybe cosmic ray) and was not selected by SDSS. But this quasar was also independently discovered by the Pan-STARRS1 high redshift quasar survey (E. Ba\~nados et al. 2016, in preparation).
The redshifts of J1545+6028 and J2356$-$0622 were determined by using ASERA to visually fit the $\rm Ly \alpha$, N\,{\sc v} emission lines, and the position of $\rm Ly \alpha$ drop.

Although we have identified four $z\sim6$ quasars with our method, we still cannot give an accurate success rate for finding $z\gtrsim5.7$ quasars as we only observed very few candidates with highest priority. We are still working on the spectroscopy observations and will also extend our method to the ongoing DECam legacy survey (DECaLS) and the unblurred coadds of the 2015 Imaging \citep[unWISE;][]{lang14} to find fainter $z\gtrsim5.7$ quasars.

\begin{figure}[tbh]
\includegraphics[width=0.5\textwidth]{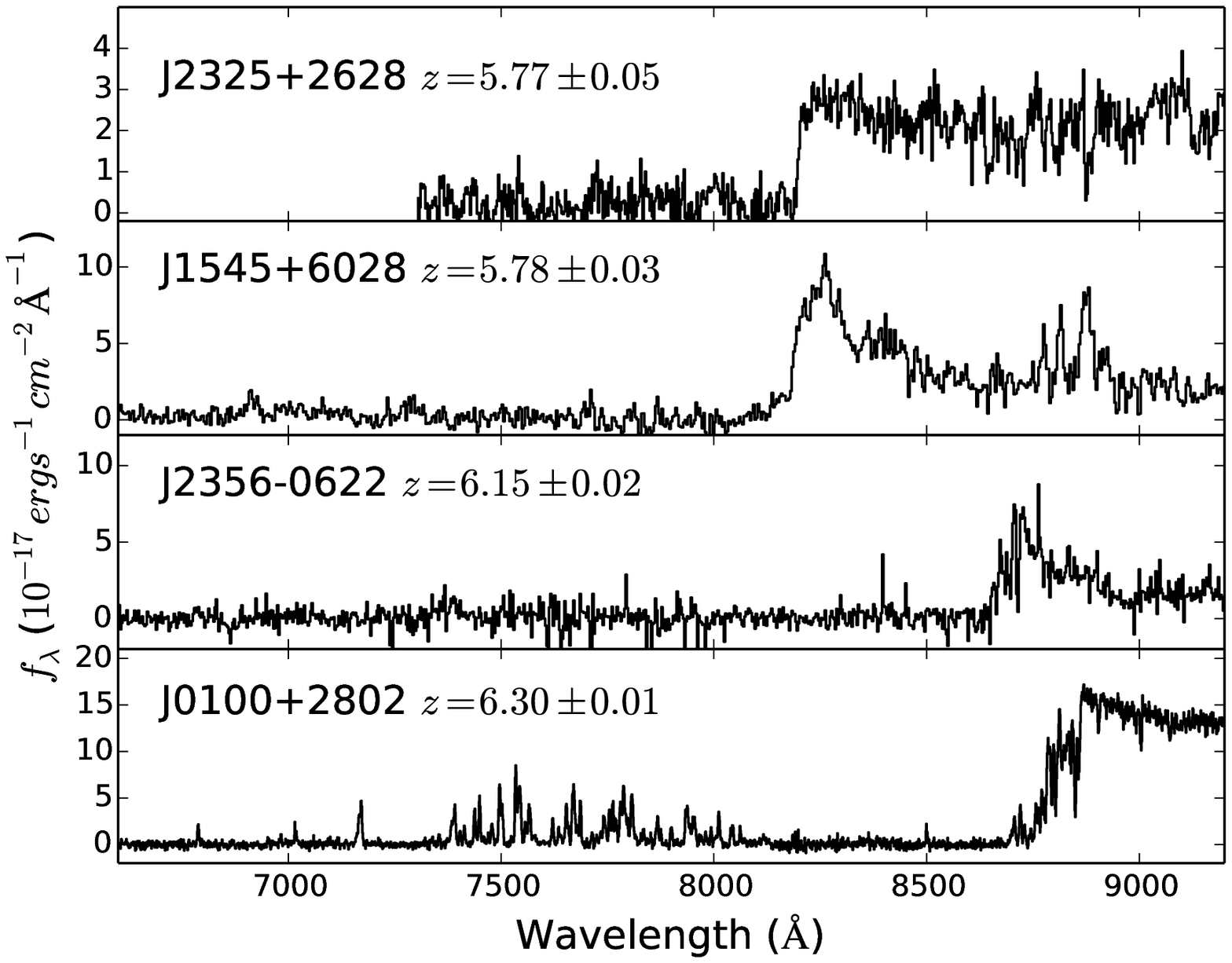}
\caption{Optical spectra of four $z\gtrsim5.7$ quasars discovered by using the method described in Section 6. J0100$+$2802 is the most luminous quasar at $z>5$ known to date and has been reported in \cite{wu15}. \label{fig6}}
\end{figure}

\section{Summary}
The SDSS spectroscopic surveys have discovered the majority of quasars known to date. However, they have a lower degree of completeness at high redshift (e.g. $z\gtrsim4.5$).  Even with additional  efforts  aimed at finding high-redshift quasars by using SDSS and other photometric data \citep[e.g.,][]{fan99,fan00b,willott07,jiang08,mortlock11,mcgreer13,venemans13}, there is an obvious gap for quasars at $5.2 < z < 5.7$. Up to now about 300,000 quasars have been spectroscopically identified, but only $\sim 700$ quasars at $z\ge4.5$, $\sim$ 30 quasars at $5.2< z < 5.7$ and $\sim$ 90 quasars at $z\ge5.7$. 

ALLWISE has a high detection rate of known high-redshift quasars, especially in the two bluest {\it WISE} bands. The $W1-W2$ color can be used to reject late-type stars efficiently. We have developed a new method by combining optical and {\it WISE} colors that yields a much higher selection efficiency for finding  luminous high-redshift quasars than by  using only optical colors. We have assembled a sample of $z\sim5.0$ quasar candidates using SDSS and ALLWISE photometric data and have spectroscopically identified 72 new $z\sim5.0$ quasars, with 12 quasars at $z\ge5.2$, allowing us to begin filling  the gap at $5.2< z < 5.7$. Our current spectroscopic observations have led to  an increase of $\sim$36\% of quasars at $5.2< z < 5.7$. However, this gap is still far from being filled as our method only focus on quasars at $z\lesssim 5.4$. A lot of work still needs to be done in order to find more quasars at this redshift range.
Our new quasar sample is about 0.8 mag brighter than SDSS $z\sim5$ quasars and is expected to set strong constraints on the bright end of the $z\sim5$ QLF \cite[][;Paper II]{yang16}, the massive end of the BHMF with future near-IR spectroscopy (Paper III), and the high-redshift IGM through future high resolution quasar absorption spectra. Moreover, our method can also be used to find quasars with redshifts up to $z\lesssim6.4$ and we have identified four quasars with redshifts beyond 5.7.

\acknowledgments
F. W. and X.-B. W. thank the support from the NSFC grants No.11373008 and 11533001, the Strategic Priority Research Program ''The Emergence of Cosmological Structures'' of the Chinese Academy of Sciences, Grant No. XDB09000000, and the National Key Basic Research Program of China 2014CB845700. F. W. thanks the financial support from the program of China Scholarships Council No. 201406010031 during his visiting in the University of Arizona.
X. F. and I. D. M. thank the support from the US NSF grant AST 11-07682. This research uses data obtained through the Telescope Access Program (TAP), which has been funded by the Strategic Priority Research Program "The Emergence of Cosmological Structures" (Grant No. XDB09000000), National Astronomical Observatories, Chinese Academy of Sciences, and the Special Fund for Astronomy from the Ministry of Finance. We acknowledge the support of the staff of the Lijiang 2.4m telescope. Funding for the telescope has been provided by CAS and the People's Government of Yunnan Province. We acknowledge the use of the MMT Observatory, a joint facility of the Smithsonian Institution and the University of Arizona. We acknowledge the use of the Bok telescope and ANU 2.3m telescope. We acknowledge the use of Xinglong 2.16m telescope. This work was partially supported by the Open Project Program of the Key Laboratory of Optical Astronomy, National Astronomical Observatories, Chinese Academy of Sciences.

We acknowledge the use of SDSS photometric data. Funding for SDSS-III has been provided by the Alfred P. Sloan Foundation, the Participating Institutions, the National Science Foundation, and the U.S. Department of Energy Office of Science. The SDSS-III Web site is http://www.sdss3.org/. SDSS-III is managed by the Astrophysical Research Consortium for the Participating Institutions of the SDSS-III Collaboration including the University of Arizona, the Brazilian Participation Group, Brookhaven National Laboratory, University of Cambridge, Carnegie Mellon University, University of Florida, the French Participation Group, the German Participation Group, Harvard University, the Instituto de Astrofisica de Canarias, the Michigan State/Notre Dame/JINA Participation Group, Johns Hopkins University, Lawrence Berkeley National Laboratory, Max Planck Institute for Astrophysics, Max Planck Institute for Extraterrestrial Physics, New Mexico State University, New York University, Ohio State University, Pennsylvania State University, University of Portsmouth, Princeton University, the Spanish Participation Group, University of Tokyo, University of Utah, Vanderbilt University, University of Virginia, University of Washington, and Yale University. 
This publication makes use of data products from the Wide-field Infrared Survey Explorer, which is a joint project of the University of California, Los Angeles, and the Jet Propulsion Laboratory/California Institute of Technology, and NEOWISE, which is a project of the Jet Propulsion Laboratory/California Institute of Technology. WISE and NEOWISE are funded by the National Aeronautics and Space Administration.

{\it Facilities:} \facility{Sloan (SDSS)}, \facility{WISE}, \facility{2.4m/YNAO (YFOSC)}, \facility{MMT (Red Channel spectrograph)}, \facility{2.16m/NAOC (BFOSC)}, \facility{2.3m/ANU (WiFeS)}.

\clearpage


\begin{thebibliography}{}
\bibitem[Ai et al.(2016)]{ai15} Ai, Y.~L., Wu, X.-B., Yang, J., et al.\ 2016, \aj, 151, 24 
\bibitem[Anderson et al.(2001)]{anderson01} Anderson, S.~F., Fan, X., Richards, G.~T., et al.\ 2001, \aj, 122, 503 
\bibitem[Arnaboldi et al.(2007)]{arnaboldi07} Arnaboldi, M., Neeser, M.~J., Parker, L.~C., et al.\ 2007, The Messenger, 127, 28 
\bibitem[Ba{\~n}ados et al.(2014)]{banados14} Ba{\~n}ados, E., Venemans, B.~P., Morganson, E., et al.\ 2014, \aj, 148, 14 
\bibitem[Ba{\~n}ados et al.(2015)]{banados15} Ba{\~n}ados, E., Venemans, B.~P., Morganson, E., et al.\ 2015, \apj, 804, 118 
\bibitem[Becker et al.(1995)]{becker95} Becker, R.~H., White, R.~L., \& Helfand, D.~J.\ 1995, \apj, 450, 559
\bibitem[Blain et al.(2013)]{blain13} Blain, A.~W., Assef, R., Stern, D., et al.\ 2013, \apj, 778, 113 
\bibitem[Bovy et al.(2011)]{bovy11} Bovy, J., Hennawi, J.~F., Hogg, D.~W., et al.\ 2011, \apj, 729, 141 
\bibitem[Carnall et al.(2015)]{carnall15} Carnall, A.~C., Shanks, T., Chehade, B., et al.\ 2015, \mnras, 451, L16 
\bibitem[Childress et al.(2014)]{childress14} Childress, M.~J., Vogt, F.~P.~A., Nielsen, J., \& Sharp, R.~G.\ 2014, \apss, 349, 617 
\bibitem[Chiu et al.(2005)]{chiu05} Chiu, K., Zheng, W., Schneider, D.~P., et al.\ 2005, \aj, 130, 13 
\bibitem[Cool et al.(2006)]{cool06} Cool, R.~J., Kochanek, C.~S., Eisenstein, D.~J., et al.\ 2006, \aj, 132, 823 
\bibitem[Croom et al.(2001)]{croom01} Croom, S.~M., Smith, R.~J., Boyle, B.~J., et al.\ 2001, \mnras, 322, L29 
\bibitem[Croom et al.(2004)]{croom04} Croom, S.~M., Smith, R.~J., Boyle, B.~J., et al.\ 2004, \mnras, 349, 1397 
\bibitem[Dawson et al.(2013)]{dawson13} Dawson, K.~S., Schlegel, D.~J., Ahn, C.~P., et al.\ 2013, \aj, 145, 10 
\bibitem[DiPompeo et al.(2015)]{dipompeo15} DiPompeo, M.~A., Bovy, J., Myers, A.~D., \& Lang, D.\ 2015, \mnras, 452, 3124 
\bibitem[Dopita et al.(2007)]{dopita07} Dopita, M., Hart, J., McGregor, P., et al.\ 2007, \apss, 310, 255 
\bibitem[Dopita et al.(2010)]{dopita10} Dopita, M., Rhee, J., Farage, C., et al.\ 2010, \apss, 327, 245 
\bibitem[Eisenstein et al.(2011)]{eisenstein11} Eisenstein, D.~J., Weinberg, D.~H., Agol, E., et al.\ 2011, \aj, 142, 72 
\bibitem[Fan et al.(2006a)]{fan06a} Fan, X., Carilli, C.~L., \& Keating, B.\ 2006a, \araa, 44, 415 
\bibitem[Fan et al.(2004)]{fan04} Fan, X., Hennawi, J.~F., Richards, G.~T., et al.\ 2004, \aj, 128, 515 
\bibitem[Fan et al.(2001a)]{fan01a} Fan, X., Narayanan, V.~K., Lupton, R.~H., et al.\ 2001a, \aj, 122, 2833 
\bibitem[Fan et al.(2006b)]{fan06b} Fan, X., Strauss, M.~A., Richards, G.~T., et al.\ 2006b, \aj, 131, 1203 
\bibitem[Fan et al.(2001b)]{fan01b} Fan, X., Strauss, M.~A., Richards, G.~T., et al.\ 2001b, \aj, 121, 31 
\bibitem[Fan et al.(2003)]{fan03} Fan, X., Strauss, M.~A., Schneider, D.~P., et al.\ 2003, \aj, 125, 1649 
\bibitem[Fan et al.(2000a)]{fan00a} Fan, X., Strauss, M.~A., Schneider, D.~P., et al.\ 2000a, \aj, 119, 1 
\bibitem[Fan et al.(1999)]{fan99} Fan, X., Strauss, M.~A., Schneider, D.~P., et al.\ 1999, \aj, 118, 1 
\bibitem[Fan et al.(2000b)]{fan00b} Fan, X., White, R.~L., Davis, M., et al.\ 2000b, \aj, 120, 1167 
\bibitem[Glikman et al.(2006)]{glikman06} Glikman, E., Helfand, D.~J., \& White, R.~L.\ 2006, \apj, 640, 579 
\bibitem[Jiang et al.(2008)]{jiang08} Jiang, L., Fan, X., Annis, J., et al.\ 2008, \aj, 135, 1057 
\bibitem[Jiang et al.(2009)]{jiang09} Jiang, L., Fan, X., Bian, F., et al.\ 2009, \aj, 138, 305 
\bibitem[Jiang et al.(2015)]{jiang15} Jiang, L., McGreer, I.~D., Fan, X., et al.\ 2015, \aj, 149, 188 
\bibitem[Kaiser et al.(2002)]{kaiser02} Kaiser, N., Aussel, H., Burke, B.~E., et al.\ 2002, \procspie, 4836, 154 
\bibitem[Kaiser et al.(2010)]{kaiser10} Kaiser, N., Burgett, W., Chambers, K., et al.\ 2010, \procspie, 7733, 77330E 
\bibitem[Kellermann et al.(1989)]{kellermann89} Kellermann, K.~I., Sramek, R., Schmidt, M., Shaffer, D.~B., \& Green, R.\ 1989, \aj, 98, 1195 
\bibitem[Kirkpatrick et al.(2011)]{kirkpatrick11} Kirkpatrick, J.~D., Cushing, M.~C., Gelino, C.~R., et al.\ 2011, \apjs, 197, 19 
\bibitem[Kochanek et al.(2012)]{kochanek12} Kochanek, C.~S., Eisenstein, D.~J., Cool, R.~J., et al.\ 2012, \apjs, 200, 8 
\bibitem[Lang(2014)]{lang14} Lang, D.\ 2014, \aj, 147, 108 
\bibitem[Lawrence et al.(2007)]{lawrence07} Lawrence, A., Warren, S.~J., Almaini, O., et al.\ 2007, \mnras, 379, 1599 
\bibitem[Leipski et al.(2014)]{leipski14} Leipski, C., Meisenheimer, K., Walter, F., et al.\ 2014, \apj, 785, 154 
\bibitem[Lupton et al.(1999)]{lupton99} Lupton, R.~H., Gunn, J.~E., \& Szalay, A.~S.\ 1999, \aj, 118, 1406 
\bibitem[Mainzer et al.(2011)]{mainzer11} Mainzer, A., Bauer, J., Grav, T., et al.\ 2011, \apj, 731, 53 
\bibitem[McGreer et al.(2013)]{mcgreer13} McGreer, I.~D., Jiang, L., Fan, X., et al.\ 2013, \apj, 768, 105 
\bibitem[Morganson et al.(2012)]{morganson12} Morganson, E., De Rosa, G., Decarli, R., et al.\ 2012, \aj, 143, 142 
\bibitem[Mortlock et al.(2011)]{mortlock11} Mortlock, D.~J., Warren, S.~J., Venemans, B.~P., et al.\ 2011, \nat, 474, 616 
\bibitem[P{\^a}ris et al.(2012)]{paris12} P{\^a}ris, I., Petitjean, P., Aubourg, {\'E}., et al.\ 2012, \aap, 548, A66 
\bibitem[P{\^a}ris et al.(2014)]{paris14} P{\^a}ris, I., Petitjean, P., Aubourg, {\'E}., et al.\ 2014, \aap, 563, A54 
\bibitem[Pogge et al.(2010)]{pogge10} Pogge, R.~W., Atwood, B., Brewer, D.~F., et al.\ 2010, \procspie, 7735, 77350A 
\bibitem[Richards et al.(2002)]{richards02} Richards, G.~T., Fan, X., Newberg, H.~J., et al.\ 2002, \aj, 123, 2945 
\bibitem[Richards et al.(2006)]{richards06} Richards, G.~T., Lacy, M., Storrie-Lombardi, L.~J., et al.\ 2006, \apjs, 166, 470 
\bibitem[Schmidt et al.(1989)]{schmidt89} Schmidt, G.~D., Weymann, R.~J., \& Foltz, C.~B.\ 1989, \pasp, 101, 713 
\bibitem[Schneider et al.(2001)]{schneider01} Schneider, D.~P., Fan, X., Strauss, M.~A., et al.\ 2001, \aj, 121, 1232 
\bibitem[Schneider et al.(2010)]{schneider10} Schneider, D.~P., Richards, G.~T., Hall, P.~B., et al.\ 2010, \aj, 139, 2360 
\bibitem[Shanks et al.(2015)]{shanks15} Shanks, T., Metcalfe, N., Chehade, B., et al.\ 2015, \mnras, 451, 4238 
\bibitem[Shen et al.(2011)]{shen11} Shen, Y., Richards, G.~T., Strauss, M.~A., et al.\ 2011, \apjs, 194, 45 
\bibitem[Stern et al.(2000)]{stern00} Stern, D., Spinrad, H., Eisenhardt, P., et al.\ 2000, \apjl, 533, L75 
\bibitem[Vanden Berk et al.(2001)]{vanden01} Vanden Berk, D.~E., Richards, G.~T., Bauer, A., et al.\ 2001, \aj, 122, 549 
\bibitem[Venemans et al.(2015)]{venemans15} Venemans, B.~P., Ba{\~n}ados, E., Decarli, R., et al.\ 2015, \apjl, 801, L11 
\bibitem[Venemans et al.(2013)]{venemans13} Venemans, B.~P., Findlay, J.~R., Sutherland, W.~J., et al.\ 2013, \apj, 779, 24 
\bibitem[Wang et al.(2015)]{wang15} Wang, F., Wu, X.-B., Fan, X., et al.\ 2015, \apjl, 807, L9 
\bibitem[Willott et al.(2010a)]{willott10a} Willott, C.~J., Albert, L., Arzoumanian, D., et al.\ 2010, \aj, 140, 546 
\bibitem[Willott et al.(2007)]{willott07} Willott, C.~J., Delorme, P., Omont, A., et al.\ 2007, \aj, 134, 2435 
\bibitem[Willott et al.(2010b)]{willott10b} Willott, C.~J., Delorme, P., Reyl{\'e}, C., et al.\ 2010, \aj, 139, 906 
\bibitem[Willott et al.(2009)]{willott09} Willott, C.~J., Delorme, P., Reyl{\'e}, C., et al.\ 2009, \aj, 137, 3541 
\bibitem[Wright et al.(2010)]{wright10} Wright, E.~L., Eisenhardt, P.~R.~M., Mainzer, A.~K., et al.\ 2010, \aj, 140, 1868 
\bibitem[Wu et al.(2012)]{wu12} Wu, X.-B., Hao, G., Jia, Z., Zhang, Y., \& Peng, N.\ 2012, \aj, 144, 49 
\bibitem[Wu et al.(2015)]{wu15} Wu, X.-B., Wang, F., Fan, X., et al.\ 2015, \nat, 518, 512 
\bibitem[Yang et al.(2016)]{yang16} Yang, J., Wang, F., Wu, X.-B., et al.\ 2016, \apj, submitted
\bibitem[Yi et al.(2014)]{yi14} Yi, W.-M., Wang, F., Wu, X.-B., et al.\ 2014, \apjl, 795, L29 
\bibitem[York et al.(2000)]{york00} York, D.~G., Adelman, J., Anderson, J.~E., Jr., et al.\ 2000, \aj, 120, 1579 
\bibitem[Yuan et al.(2013)]{yuan13} Yuan, H., Zhang, H., Zhang, Y., et al.\ 2013, Astronomy and Computing, 3, 65 
\bibitem[Zheng et al.(2000)]{zheng00} Zheng, W., Tsvetanov, Z.~I., Schneider, D.~P., et al.\ 2000, \aj, 120, 1607 

\end{thebibliography}
\end{document}